\documentclass{aastex}
\usepackage{epic}

\shorttitle{Corotation resonances}
\shortauthors{Vera-Villamizar et al.}

\begin{document}

\title{ANALYSIS OF RESONANCES IN GRAND DESIGN SPIRAL GALAXIES}

\author{Nelson Vera-Villamizar and Horacio Dottori}
\affil{Instituto de F\'{\i}sica -- Universidade Federal do Rio Grande
do Sul,\\ CxP 15051, CEP 91501-970, Porto Alegre, RS, Brazil}

\author{Iv\^anio Puerari}
\affil{Instituto Nacional de Astrof\'{\i}sica, Optica y Electr\'onica,\\
Calle Luis Enrique Erro 1, 72840 Tonantzintla, Puebla, M\'exico}

\and 

\author{Reinaldo de Carvalho}
\affil{Observat\'orio Nacional, Rua General Jos\'e Cristino 77, \\
CEP 20921-400, Rio de Janeiro, RJ, Brazil}

\begin{abstract}

We have searched for corotations (CR) in three Southern grand design 
spiral galaxies: NGC\,1365, NGC\,1566 and NGC\,2997. We have 
also introduced a method of quantifying errors in the phase 
diagram used to detect CR. 

We established the m=2 pattern CR at 12.1\,kpc, 9.4\,kpc and 
7\,kpc, for NGC\,1365, NGC\,1566 and NGC\,2997, respectively.
By using published rotation curves, we could determine
spiral pattern angular speeds of 25.0\,km/sec/kpc, 
12.2\,km/sec/kpc and 17.6\,km/sec/kpc, respectively.

A three armed component has been detected in 
NGC\,2997, with the CR placed at 8.7\,kpc with a
pattern angular speed $\Omega_{CR_3}=12.7$\,km/sec/kpc. 

An m=1 component was detected in NGC\,1566. We warily locate 
the CR at 7.1\,kpc, with a pattern angular speed
$\Omega_{CR}\backsimeq16.6$\,km/sec/kpc. This pattern 
does not present ILR.

Ages have been determined by studying the radial density 
profile of the m=2 Fourier components in $g$ (newly formed 
stars) and $i$ (perturbing SDW supported by the disk of old 
stars), aided by the global aspect of the real spiral pattern 
in comparison with numerical simulations. The pattern is 
$\thicksim$1200\,Myr old in NGC\,1365, $\thicksim$800\,Myr old 
in NGC\,1566 and younger than 80\,Myr in NGC\,2997. 

\end{abstract}


\section{Introduction}

Within the framework of the Density Wave Theory \citep{linshu64}, 
the spiral arms of disk galaxies are the manifestation of traveling
waves. In a first approximation, the pattern speed of a density
wave is constant over the disk. It is widely known that  galactic
disks present differential rotation and in this way, in a given 
galaxy, there is a radial position where the density wave pattern 
speed coincides with the angular velocity of the stellar disk. 
This radial position is know as the corotation resonance (CR) 
radius. The density wave pattern is slower than the disk material 
at radii smaller than the CR, and it is faster at radii larger than
the CR, inducing shocks at different arms side inward and outward
the CR. Other resonances can be present in a given galaxy, mainly
the ILR (Inner Lindblad Resonance), and the OLR (Outer Lindblad
Resonance). These resonances are placed when the condition
$\Omega_p=\Omega(R)\pm\kappa/2$ is satisfied. Here, $\Omega_p$
is the pattern speed, $\Omega(R)$ is the angular velocity of the 
disk, $\kappa$ is the epicyclic frequency, and the signals `$+$' and 
`$-$' refer to the ILR and the OLR, respectively.

\cite{schweizer76} and \cite{bece90} have previously discussed 
what would be the behavior of the colors across spiral arms if a 
shock generated by a spiral density wave (SDW) induces star formation.
The main observable characteristic of this scenario are 
steeper azimuthal profiles and bluer color indices on the side 
where the shock front is located.

\cite{puedot97} proposed a photometric method to detect CR based on
the comparison of blue and infrared frames azimuthal profiles Fourier 
transform, under the basic idea that a shock-induced star formation 
in a SDW scenario produces an azimuthal gradient 
of ages across the spiral arms that has {\it opposite signs} on 
either side of the CR.

The behavior of the azimuthal phase {$\Theta(r)$} was schematically 
exemplified in Fig.\,2 of \cite{puedot97}, for leading and trailing 
waves of S- and Z-type. \cite{eem92} pointed out that evidences for 
CR are clearly seen in gas rich galaxies in the form of sharp end points 
to star formation ridges and dust lanes in two armed spirals. 
Photometric methods are widely used to detect resonances, mainly 
due to the spar in observing time. \cite{canzian98} rise some 
doubts on the ability of photometric methods to uniquely determine 
CR in M\,100. For this galaxy, \cite{ees89}, using computing 
enhanced imagery, placed the CR radius at 110\,arcsec. This value 
was confirmed by \cite{sempetal95}, through the HI velocity field.
On the other hand, two different methods based on H$\alpha$ velocity
field place the CR of M\,100 at around 75\,arcsec 
\citep{arsenaultetal88,canzianallen97}.
That shows for M\,100 that even the kinematical determination of 
CR from the velocity field of two different components do not 
lead to a unique result. Perhaps, the HI streaming motions over 
M\,100 arms \citep{knapenetal92} contribute to disturb SDW 
kinematics, indicating that perturbations in disks of galaxies 
sometimes produce more complex spatial structures and/or
kinematical patterns than those suggested by simple models.

In this paper, we use and discuss again the azimuthal profiles 
phase method \citep{puedot97} to analyze the grand design spiral
galaxies NGC\,1365, NGC\,1566 and NGC\,2997, recalling the 
curious result on multiple CR obtained by these authors for 
NGC\,1832 and NGC\,7479.
NGC\,1365 and NGC\,1566 belong to \cite{elmelm87} arm class 12
(`two long symmetric arms dominating the optical disk'), while 
NGC\,2997 belongs to the arm class 9 (`two symmetric inner arms; 
multiple long and continuous outer arms').

Furthermore, in this 
paper we present a method to quantitatively determine errors. 
This procedure allows to determine a range of validity in radius 
for the azimuthal profiles phase method, a critical issue in 
\cite{puedot97} Letter. We apply the azimuthal profiles phase 
method on the rough $g$ and $i$ images, as done by \cite{puedot97}, 
but also to images computer-treated with the  method of 
\cite{eem92}, and that based on modes search by two-dimensional 
Fourier analysis \citep{kalnajs75,consathan82,puedot92}.

\section{Data and Analysis}

\subsection{Observations}

The galaxies discussed in this paper were observed in November, 
1998, at the CTIO 0.9m telescope. For each galaxy, three images 
with each of Gunn's $g$ and $i$ filters were taken, with exposure 
time 500 seconds each image. The standard reduction was carried
out using the IRAF package\footnote{The IRAF package is written 
and supported by the IRAF programming group at the National 
Optical Astronomy Observatories (NOAO) in Tucson, Arizona. NOAO 
is operated by the Association of Universities for Research in 
Astronomy (AURA), Inc. under cooperative agreement with the 
National Science Foundation (NSF).}. Since our focus is morphology, 
no calibration were required. As a first step, all images were 
centered using field stars. Thereafter, field stars were removed 
using the IRAF {\it imedit} task. In order to enhance the 
perturbation over the galaxy's disks we subtract the mean radial 
light profile and then we normalize the rms variation of the 
intensity at each radius to a constant value throughout the image, 
as done by \cite{eem92}. The galaxies were then deprojected 
running a program that maps ellipses into circles.

\subsection{The Azimuthal Profile Phase Method}

Shock induced star formation in a stellar density wave scenario 
produces an azimuthal spread of ages across the spiral arms. At 
the CR, the angular velocity of the perturbation ($\Omega_p$) 
and that of the stellar disk ($\Omega$) coincide. A comoving 
observer at the CR will see outward and inward the shock front 
to change from one side of the spiral to the other, consequently 
reversing the order in which young and older disk stellar 
population appear in azimuthal profiles across the arms. In order 
to detect the shock front jump, \cite{puedot97} proposed to 
analyze the relative behavior of the SDW and shock front phases 
respectively, by means of the Fourier transform of azimuthal 
profiles $I_r(\theta)$ given for m=2 by

$$F_2(r)\,=\,\int_{-\pi}^{+\pi}I_r(\theta)e^{-2i\pi}d\theta$$

The phase $\Theta(r)$ can be obtained for  m=2 as

$$\Theta_2(r)\,=\,tan^{-1}\frac{Re[F_2(r)]}{Im[F_2(r)]}$$

\noindent where $Re$ and $Im$ mean the real and imaginary parts
of the complex Fourier coefficients. In Figs. 1 and 2 of 
\cite{puedot97}, one can find a graphic representation of
$\Theta(r)$ behavior for `S' and `Z' arms of trailing and leading 
character. This method was also successfully used by \cite{abp98} to
find CR in barred galaxies.

The SDW phase and the newly formed stars one are respectively 
identified with $\Theta_{dw_2}$\,=\,$\Theta_{2i}$ and 
$\Theta_{sf_2}$\,=\,$\Theta_{2g}$, where the right 
terms are obtained from infrared $i$ and blue $g$ images,
respectively.
An Azimuthal Profile Phase Difference (hereinafter, APPD)
$[\Theta_{2g}- \Theta_{2i}]=0$, will indicate the presence of a CR.

We also search in this paper for the existence of the m=3 
component. For this component the phase $\Theta$ is given by 

$$\Theta_3(r)\,=\,tan^{-1}\frac{Re[F_3(r)]}{Im[F_3(r)]}$$

\noindent where

$$F_3(r)\,=\,\int_{-\pi}^{+\pi}I_r(\theta)e^{-3i\pi}d\theta$$

\subsection{The Errors}

The basic idea for errors determination is that if the phase
differences in the comparison of $g$ and $i$ images reflect an 
statistical fluctuation, it should also be reflected when 
different single filter images are compared among them.
In order to quantitatively evaluate the errors intrinsic to 
the APPD method, we obtained three different images in
$g$ and $i$ filters. We co-added $g1$+$g2$ and $g2$+$g3$ and compute 
the phase difference of the resulting images. The same procedure 
is applied to the $i$ images. When comparing single filter images 
of a given galaxy, the global aspect of the APPD is that of a high
frequency statistical fluctuation (white noise), while APPD of $g$ 
against $i$ images produces low frequency phase differences, as we 
will see further on. We assume that an APPD that locally reach values 
larger than 3$\sigma$ constitutes a real signal.

\subsection{Separation of m=2 and m=3 Components}

In order to separate m=2 and m=3 spiral pattern components we apply
two methods, namely, that of \cite{eem92} and the one based on 
two-dimensional Fourier transform \citep{puedot92}.

\subsubsection{Elmegreen, Elmegreen \& Montenegro Method (EEM92)}

These authors propose to separate the two- and three-fold symmetric
part of a galaxy spiral pattern, S$_2$ and S$_3$ respectively, by 
making  successive images rotations and subtractions. If 
$I(r, \theta)$ is the original image in polar coordinates, then

$$S_2(r, \theta)\,=\,I(r, \theta)\,-[I(r, \theta)\,-\,
I(r, \theta+\pi)]_T$$

\noindent and

$$S_3(r, \theta)\,=\,2I(r,\theta)\,-\,[I(r,\theta)\,+\,I(r,\theta+
\frac{2\pi}{3})]_T\,-\,[I(r, \theta)\,-\,I(r,\theta-\frac{2\pi}{3})]_T$$

\noindent where $T$ stands for truncation, meaning that pixels with
negative intensities are set to 0.

\subsubsection{The Fourier method}

This method has been extensively discussed in a number of papers
\citep{kalnajs75,consathan82,iyeetal82,puedot92, puerari93}.
In the Fourier method, an image is decomposed on a basis of 
logarithmic spiral of the form $r=r_o {\rm exp} (-{m\over p} 
\theta)$. The Fourier coefficients $A(p,m)$ can be written as

$$ A(p,m) = \frac{1}{D}\int_{-\pi}^{+\pi}\int_{-\infty}^{+\infty}
I(u,\theta) {\rm exp}[-i(m \theta + p u)] d u d \theta$$

Here, $u \equiv {\rm ln}\;r$, $r$ and $\theta$ are the polar 
coordinates, $m$ represents the number of the arms, $p$ is 
related to the pitch angle $\alpha$ of the spiral as $\alpha={\rm atan}
(-m/p)$, and $I(u,\theta)$ is the distribution of light of a 
given deprojected galaxy, in a \hbox{${\rm ln}\;r$ {\sl versus} 
$\theta$ } plane. $D$ is a normalization factor written as

$$ D=\int_{-\pi}^{+\pi} \int_{-\infty}^{+\infty} I(u, \theta) du
d\theta $$

\noindent Obs.: In practice, the integrals in $u \equiv {\rm ln}\;r$ 
are calculated from a minimum radius (selected to exclude the bulge 
where there is no information on the arms) to a maximum radius 
(which extends to the outer limits of the arms in our images).

The inverse Fourier transform can be written as

$$T(u,\theta) = \sum_m P_m (u) {\rm e}^{im \theta} $$

\noindent where $P_m$ stands for the radial density profile of the $m$
component, written as

$$ P_m(u) = \frac{D}{{\rm e}^{2u} 4 \pi^2} \int_{p_-}^
{p_+}\;G_m (p) A(p,m) {\rm e}^{i p u}\;dp$$

\noindent and $G_m(p)$ is a high frequency filter used to smooth 
the $A(p,m)$ spectra at the interval ends \citep{puedot92}, 
which has the form

$$ G_m(p) = {\rm exp} \left[ -\frac{1}{2} \left( \frac{p - p_{max}^m}{25}
\right)^2 \right] $$

\noindent where $p_{max}^m$ is the value of $p$ for which the 
amplitude of the Fourier coefficients for a given $m$ is maximum. 
The chosen interval ends ($p_+=50$ and $p_-=-50$), as well as 
the step $dp=0.25$, are suitable for the analysis of galactic 
spiral arms.

So, by using 2D Fourier transforms, we can separate any m-mode. 
For example, the m=2 mode can be analyzed using $S_2(u,\theta)=
S_2(u){\rm e}^{i2\theta}$, and so on.

\section{Spiral Structure Properties in NGC\,1365}

The disk subtracted and rectified to face-on images of NGC\,1365 
in the $g$ and $i$ filters, are presented in Figure \ref{ima1365_g_i}.
Two dimensional Fourier analysis shows that the spiral structure in
NGC\,1365 is mainly composed of  m=2 components, in both, $g$ and $i$
frames (Figure \ref{fouspec1365}). The strong signal appearing in m=4 
and m=6 are only aliases of m=2 \citep{puedot92}. Nevertheless, 
the radial dependence of the m=2 arm intensity P$_2$(r) is not 
similar in both frames,  as Figure \ref{smr1365} shows. Up to 10\,kpc, 
the SDW intensity, represented by the old stellar component, 
P$_{2i}(r)$, predominates on that of the perturbed material,
represented by the newly formed stars, P$_{2g}(r)$. From 13\,kpc 
outwards, P$_{2i}(r)$ weakens compared to P$_{2g}(r)$.
When compared with numerical models, it becomes an interesting 
method to determine the pattern age, as we will see in section 3.2.

\subsection{CR Resonance in NGC\,1365}

The identification of resonances in barred spirals is a matter of 
increasing interest for many reasons related to the SDW theory.  
By example, the  self-consistent modeling of real systems 
\cite{Lind2_96},  the study of the time evolution of SDW (formation, 
dumping, etc) and related induced star formation 
\citep{juncom96,sell_spar88,rau_sa99}, and the study of non-linear 
coupling among different spirals modes 
\citep{Tagger87, Sygnet88, mas_ta_97b}, are interesting, even open, 
problems. 

Resonances in NGC\,1365 have been  kinematically analyzed by
\cite{Jors95} and \cite{Lind2_96}. They place the CR resonance at 
13.8\,kpc and 15.2\,kpc (R$_{CR}$/R$_{bar}$\, 1.15 and 1.3), 
respectively. \cite{Lind2_96}, point that in models where the 
disk suffers a pure bar perturbation, a pattern speed higher 
than 20\,$\pm$\,1\,km/sec/kpc (corresponding to smaller CR radii) 
tends to give spirals more tighty wound than observed. They also 
point out that the CR radius should be even more external, if the 
perturbation is composed by a bar plus spiral arms. Nevertheless, as 
mentioned by these authors the evolution of the gaseous and/or 
stellar system to its present form is beyond the scope of their 
paper. This evolution precisely transforms dramatically 
the spiral pattern  parameters (number of arms, pitch angle, etc.), 
distinctly in old and young components, as can be seen in 
\cite{juncom96} self-consistent numerical experiments. This might 
explain the difference between \cite{Jors95} and \cite{Lind2_96} 
results. Our results are more similar to those of \cite{Jors95}, 
as we will see in the next paragraph.

In  Figure \ref{rectifica1365} we present the rectified $g$ image, 
in a log\,r vs. $\theta$ diagram \citep{eem92}. In this plot we can 
see more precisely the different arm structures. The bar extends up 
to 10.3\,kpc and its axis is strictly straight, as can be better 
seen in the western arm, which is less disturbed by the dust lane. 
The dust lane presents an open spiral form, with pitch angle
$\alpha\,=\,67^\circ$ with extension similar to that of the bar. 
The bisymmetrical arms present two different pitch angles. At 
the beginning it has $\alpha\,=\,16^\circ$, and extends itself from 
the end of the bar up to 17.5\,kpc. The rest of the arm, up to 
$\approx$\,28\,kpc, has $\alpha\,=\,40^\circ$. It has to be pointed 
out that the pitch angle change is not an artifact introduced 
by the galaxy warping, since the galaxy was corrected by an 
inclination $\omega\,=\,40^\circ$ and the warping does not depart 
more than 15$^\circ$ from this value. Moreover, the warping 
begins at $\approx$25\,kpc \citep{Jors95}. 

We present in Figures 
\ref{phase+error_pure1365}, \ref{phase+error_eem1365} and
\ref{phase+error_four1365} the behavior of the m=2 APPD 
of the original images ($[\Theta_{2g}\,-\,\Theta_{2i}]_{orig}$), 
and those transformed by EEM92 
($[\Theta_{2g}\,-\,\Theta_{2i}]_{eem}$) and Fourier 
($[\Theta_{2g}\,-\,\Theta_{2i}]_{Four}$) methods. 
The corresponding errors in $g$ and $i$, are obtained according 
to the prescriptions of Section\,2.3. 

In Figure \ref{phase+error_pure1365}, one can  see that inside 
5\,kpc the S/N is too high, mainly in the $i$ band, as to advance 
any  conclusion. Between $\approx$ 6\,kpc and 15\,kpc,
$[\Theta_{2g}\,-\,\Theta_{2i}]$ comfortably  overcome 
the noise by many $\sigma$. Beyond 15\,kpc, it is again strongly 
affected by the noise. The CR at the extreme of the bar is set 
at 10.7\,kpc by this method. The more internal cuts are probably 
produced by the influence of the dust lane, whose curvature is 
different from that of the stellar bar.

$[\Theta_{sf}\,-\,\Theta_{dw}]_{eem}$ and
$[\Theta_{sf}\,-\,\Theta_{dw}]_{Four}$  present a similar behavior 
than that of the original images, but with a better S/N. The CR is set at 13.8 
and 11.8\,kpc respectively by these two methods, with a mean value 
in close agreement to \cite{Jors95} determination. Both methods 
show a tendency to unravel a CR at 25\,kpc, distance at which the 
disk begin to warp.  But the error in the $i$ image increase
strongly, and prevent any conclusion on the reality of this CR.
Deeper imagery could help to better understand the SDW behavior
in this region.

According to the rotation curve of \cite{per_sal95},
the mean value for the CR radius of r$_{CR}$=12.1\,kpc leads to a 
pattern speed $\Omega_{Pattern}$=25.0\,Km/s/kpc.

We have searched for the presence 
of m\,=\,3 component, by means of EEM92 and Fourier transform 
methods, but this component is not significant in NGC\,1365
(see the spectrum for m=3 in Fig. \ref{fouspec1365}). The m=1 
component appears to weak in the Fourier analysis, and prevent
any CR study.

\subsection{NGC\,1365 Spiral Pattern Age}

\cite{juncom96} performed numerical models of disk galaxies with  
n\,=\,1 Toomre's disks of stars and gas. In all the experiments they 
used a radial scale length, a$_s$\,=\,3.5\,kpc and 4.0\,kpc, for 
the star disk, and a$_g$\,=\,6.0\,kpc for the gaseous disk. The 
extend of their stellar and gaseous disks is 12\,kpc and  16\,kpc, 
respectively. Their models were completed with a Plummer's bulge 
with scale length, a$_b$ ranging from 1/4.5 to 1/3 of a$_s$, and 
mass equal to that of the stellar disk. These models do not have 
halo. \cite{juncom96} present the evolution of the stars and gas 
spiral patterns between 20\,Myr and 2000\,Myr for ten different 
experiments. NGC\,1365 compares very well to their experiment E. 
It has a$_s$\,=\,3 kpc, and a$_b$\,=\,1. In this numerical experiment, 
they found that the arm structure, which is multiple armed at 
200\,Myr, becomes two-armed at 800\,Myr. From there  on, the 
radial behavior of the m=2 Fourier intensity of young P$_{2g}$
and old P$_{2i}$ stellar population, which were fairly similar 
at 200\,Myr, begin to differentiate between them. P$_{2i}$
predominates in the inner 1/4 of the disk length, and P$_{2g}$
outwards. We compare  P$_{2g}$ and P$_{2i}$ in Figure 
\ref{smr1365}, where we plotted corresponding Fourier radial 
intensities for perturbing and perturbed material \citep{juncom96}, 
after 1200\,Myr (their Figure\,10). As can be seen, the qualitative 
behavior is quite similar to that of NGC\,1365, giving to this 
diagram a good chance as a model dependent spiral pattern age 
indicator. These experiments are specially suited for NGC\,1365 
because this galaxy does not possess  an important halo, as clearly 
demonstrated by the highly Keplerian behavior of the rotation 
curve for r$\,>\,\approx$\,20\,kpc \citep{Jors95}.

\section{Spiral Structure Properties of NGC\,1566}

The images in the $g$ and $i$ filters for NGC\,1566, disk subtracted 
and rectified to face-on, are presented in Figure 
\ref{ima1566_g_i}.

The rotation curve of this galaxy \citep{per_sal95} presents
a high dispersion, because it is  projected practically 
face-on on the sky . For this reason, a photometric analysis 
of the resonances is very appropriated in the case of NGC\,1566 
\citep{el_el90}.

The HII regions spiral pattern  of this galaxy was already
analyzed using Fourier transform by \cite{pue_dot90}. The 
predominant pattern of this young stellar population is m\,=\,2, 
with m\,=\,1 being progressively more important at radii 
larger than 17\,kpc. The present analysis of $g$ and $i$ imagery 
shows up the same property (Figure \ref{fouspec1566}). Contrary 
to the case of NGC\,1365, in NGC\,1566 the m\,=\,2 SDW intensity 
in the old stellar component P$_{2i}$ does not predominates 
in any part of the disk on that of the  newly formed stars, 
P$_{2g}$ (Figure \ref{smr1566}). Both functions decrease 
similarly and fade up simultaneously at about 22\,kpc. 

For the one armed component, we note that between 10 and 13\,kpc 
P$_{1i}$ dominates on P$_{1g}$, while in more external 
regions the contrary happens (Figure \ref{sm=1r1566}).

\subsection{CR Resonance in NGC\,1566}

The resonances of NGC\,1566 were studied by \cite{el_el90}. 
The method they used was an identification of optical features 
together with constrains set by plausible rotation curves. 
They fitted radii for the ILR, OLR, CR, and other resonances.

As in the case of NGC\,1365, in Figures \ref{phase+error_pure1566}, 
\ref{phase+error_eem1566}, \ref{phase+error_four1566} we present 
the behavior of the APPD for the three methods used in this 
analysis. As before, the corresponding errors in $g$ and $i$, are 
obtained according to the prescriptions of Section\,2.3. 

Each one of the methods respectively places the CR at R$\approx$ 
9.7\,kpc (Fig. \ref{phase+error_pure1566}), R$\approx$ 8.5\,kpc 
(Fig. \ref{phase+error_eem1566}), and R$\approx$ 10\,kpc (Fig.
\ref{phase+error_four1566}) in good agreement among them. 
This distance corresponds to the CR assumed by \cite{el_el90}. 
While some other intersections can be seen in the diagrams, 
the S/N indicates that their evidence is marginal. 

In Fig. \ref{rectifica1566} we present the rectified image in a
log\,r vs. $\theta$ diagram. As we can see, at the CR circle  
the main arms begin to thick outwards. The same 
behavior can be seen in the log\,r vs. $\theta$ diagram of the 
symmetrized S2 image (not presented here). Other important property 
of this galaxy is that the arms, although ill defined between 10 
and 15\,kpc, are logarithmic between 5 and 22\,kpc, with a pitch 
angle $\alpha\simeq 36^\circ$. A short secondary arm, parallel to the 
main one, appears between 12 and 22\,kpc. It is better defined in 
the SW arm, appearing as a rather asymmetric structure. Surprisingly 
a similar structure appears at 800 Myr in Junqueira and Combes 
experiment F2 (their Fig. 19), in agreement with the age determined 
in the next section for the SDW in this galaxy. The 
arms bend themselves between 20 and 24\,kpc, with a different 
pitch angle, although the feature is too weak to further analysis. 
We search also for m=3 component in NGC\,1566, but it is not 
important. On the other hand, an m=1 component is present, although
tenuously defined. One can warily locate its CR at 7.1\,kpc, without
ILR, and pattern angular speed $\Omega_{CR}$\,=\,16.6\,km/sec/kpc.

\subsection{NGC\,1566 Spiral Pattern Age}

As in the case of NGC\,1365, we compared the spiral pattern 
properties with \cite{juncom96} models. Odd SDW components in 
\cite{juncom96} models appear at earlier evolutionary stages 
of the perturbation. Later on, the model spiral structure transforms 
itself into a two armed one. The more suitable model 
for the case of NGC\,1566 is that of their experiment F, with  
n\,=\,1 Toomre's disks of stars and gas, and radial scale length 
of the star and gaseous disks, a$_s$\,=\,4.0\,kpc, and 
a$_g$\,=\,6.0\,kpc, respectively. The extend of the stellar 
and gaseous disks is 12\,kpc and 16\,kpc, respectively. The 
Plummer's bulge scale length for this model is  a$_b$\,=\,a$_s$\,/4, 
and the bulge mass is equal to that of the stellar disk. 

The behavior of P$_{2g}$ and P$_{2i}$ (Figure \ref{smr1566}) 
as well as the Fourier images of old and young stars are compatible 
with an SDW age of about 800\,Myr in this galaxy. We do not know how 
important the halo is in this galaxy, in order to evaluate if 
the models are so suitable as in the former case. Nevertheless, 
Parkes 21 cm observations \citep{becker88} indicate that the 
HI extends up to 90\% of the optical disk, pointing also in this 
case to a not too much extended halo.

\section{Spiral Structure Properties of NGC\,2997}

We present in Figure \ref{ima2997_g_i} the disk subtracted and
deprojected to face-on image of NGC2997.

This galaxy is not as symmetric as NGC\,1365 and NGC\,1566. As
discussed before, NGC\,2997 is classified by \cite{elmelm87} as
arm class 9 (`two symmetric inner arms; multiple long and 
continuous outer arms'), while NGC\,1365 and NGC\,1566 are both 
arm class 12 (`two long symmetric arms dominating the optical disk'). 

Fourier analysis of HII regions was performed by \cite{consathan82}, 
\cite{puedot92} and \cite{garciaathan93}. All studies show a 
predominance of m=2 and m=3 modes. At some radii, m=3 is even stronger 
than m=2. Near-infrared images of NGC\,2997 \citep{blockpuerari99}, 
reveal a strong asymmetry on the main inner arms, suggesting the
presence of an m=1 mode, besides the m=3 mode, in the older disk 
population \citep{blocketal94}. The Fourier bisymmetric component in a 
K$'$ image shows a pitch angle $\alpha=25^{\circ}$ .

In Figure \ref{fouspec2997} we present the Fourier spectra for
the $g$ and $i$ images of NGC\,2997. The m=2 mode is, as expected, the 
predominant one. It  is important to note that the m=1  component, 
which appears at the noise level in the $g$ image, presents a 
reasonable weight with respect to m=3 in $i$ image, although 
weaker than m=3. Since in K$'$, m=1 dominates on m=3 
\citep{blockpuerari99}, and m=3 has practically a color 
independent weight relative to m=2, it is strongly suggesting 
that the m=1 component is the only one really affected 
by dust obscuration.

\subsection{CR Resonances in NGC\,2997}

It was very difficult to apply the Phase method to the original
images of NGC\,2997. The m=2 phase did follow the internal arms,
but becomes very chaotic at the distance of  the arm
bifurcation. The presence of several strong interarm features
also contributes to increase the noise and makes the results less 
reliable. So, the  methods that involve image processing, namely EEM92 
and Fourier transform, are more suitable in the case of NGC\,2997. 

To treat properly the main m=2 component,  the part of the 
bisymmetric image in Figure \ref{s2innerouter2997} containing 
the secondary  m=2 component in the E-W direction, from 7\,kpc 
outwards, was erased. Figure \ref{phase+error_eem2997in} shows 
the APPD for the main m=2 component. There is a clear cut at 
7\,kpc, with very good S/N. The CR is located where the EW 
extension of the m=2 arms appears (Figure \ref{s2innerouter2997}). 
This can be easily seen in Figure \ref{rectifica2997}, where we 
plotted the symmetrized image in $\log$ r versus $\theta$. From 
the rotation curve of \cite{sperandio95}, we derived a 2-arms SDW 
angular velocity of $\Omega_{SDW2}$\,=\,18\,km/s/kpc. The same 
place for the m=2 CR position is found from Fourier analysis
(Figure \ref{phase+error_four2997}).

EEM92 method clearly reveals in this case the 
presence of a 3-arms component (Figure \ref{s3_2997}), each 
arm covering an arch of $\approx$\,180$^\circ$ between 
3\,kpc\,$\lesssim$\,r\,$\lesssim$\,22\,kpc. Similar 3-arms morphology
has been detected by EEM92 in NGC\,628,  NGC\,1232, NGC\,5457 and NGC\,6912. 
The global sense of winding of this component is the same as that of the 
2-armed one. Nevertheless, in the case of NGC\,2997 the 3-arms are too wide 
at the beginning and present in this region a remarkable  concavity of 
leading nature between  7\,kpc\,$\lesssim$\,r\,$\lesssim$\,11\,kpc.
We dettached the concavity  with a filled curve in Figure \ref{s3_2997}. 
The blow-up in the figure shows details of the ridge of one of the 3-arms concavity.  
APPD diagram   (Figure \ref{phase+error_eem2997s3}), 
shows a intersection  at 8.5\,kpc, where the phase shift crosses downward, 
in agreement with the leading character of the described 3-arms concavity, 
indicating that star formation occurs at the ridge crests.  The detection 
of the elusive 3-armed SDW CR resonance in this galaxy is one 
important  point of this study. Nevertheless, a better understanding 
would be necessary of what the phase diagram should do for a 3-armed 
component.

From the rotation curve of \cite{sperandio95} we 
obtained a pattern speed for the 3-arms SDW 
$\Omega_{SDW3}\approx$\,13.7\,km/s/kpc. 

In Figure \ref{rectifica2997} we present NGC\,2997 rectified image 
of the m=2 component as derived from EEM92 method, in a log\,r vs 
$\theta$ diagram. The bifurcation of the m=2 component begins at 
the CR radius (r$_{cr}$\,=\,7\,kpc). The main m=2 spiral extends 
up to r=16\,kpc and the secondary between 7\,kpc and 21\,kpc. 
Both spirals are logarithmic, with pitch angle $\alpha=15^\circ$.7 and 
P=39$^\circ$.4, respectively.

\subsection{NGC\,2997 Spiral Pattern Age}

We compare the images and m=2 Fourier density profiles of NGC\,2997 
(see Figure \ref{smr2997}) with \cite{juncom96} models. Here, 
the presence of asymmetries and higher 
m modes betrays the young age of the perturbation in NGC\,2997. 
As discussed above, odd SDW components in \cite{juncom96} 
models appear at earlier evolutionary stages of the 
perturbation. In this case, it is difficult to assign an 
specific model from \cite{juncom96} which fits NGC\,2997. 
Models E, F, G, and H show similar features at stages younger 
than 80 Myr. The m=2 Fourier radial density profiles  present two 
differentiated parts: Inside 14\,kpc P$_{2g}$\,$\approx$\,
P$_{2i}$, while outwards the old population function predominates 
on that of the young one, probably indicating that the external 
parts of the arms may have been produced by an older SDW. This last assessment 
has to be taken with caution, since \cite{juncom96} model F 
that fits this galaxy,  seems to be oversimplified  to explain the 
scenario of the SDW in NGC\,2997.

\section{Discussion and Conclusions}

The one-D Fourier transform method, applied to $g$ (blue) and $i$ 
(infrared) images proposed by \cite{puedot97} to detect CR, has 
been applied to three grand design galaxies, namely, NGC\,1365, 
NGC\,1566 and NGC\,2997. Some improvements have been 
introduced in this 
paper, to make the analysis more reliable. The original 
treatment on rough images of \cite{puedot97}, was this time 
complemented with the analysis of computer treated images, with 
2-D Fourier transform and EEM92 methods. This complete
procedure presents complementary aspects, which gives advantages 
with respect to the original one. For example, to work on the 
original image is very good for pure m=2 arms. When other 
components appear, Fourier transform and EEM92 methods 
work better. Fourier method gives smaller errors than EEM92 
method, but EEM92 method is more realistic than the former, since it 
is not constrained to describe the spiral as pure logarithmic 
ones. EEM92 method also have detected nicely the presence of the m=3 
component in NGC\,2997. This method is extremely efficient 
in detecting this component, while Fourier methods has to add 
many components. So, we confident can say that m=3 is not 
present in NGC\,1365 and NGC\,1566, and also that NGC\,1566
presents an m=1 component.

The method presented here to determine errors in the APPD
diagram, also represents a considerable 
improvement with respect to the original paper \citep{puedot97}, 
that strongly helps to identify which is the level of confidence 
in a CR determination. In turn, it also helps to establish the
correct observational time to verify the dubious cases. From 
this point of view, it will be important to check \cite{puedot97} 
result on multiple CR in the galaxies NGC\,1832 and NGC\,7479. 
Nevertheless, more images of these galaxies are necessary to 
obtain a S/N in APPD diagram. The present form of this diagram, 
as a phase difference rather than as the superposition of the 
phases in each color, is more instructive than the former one, 
specially to visualize the S/N.

The determination of the arms extend has been made via log\,r vs. 
$\theta$ diagram and via rotation curve, once know the CR radius.

The arm complexity as well as the relative arm intensity for 
perturbing SDW P$_{2i}$(r) and perturbed material P$_{2g}$(r) 
in the Fourier diagram are related to the perturbations age. 
Although model dependent, a comparison with suitable models, 
allows to establish a scale of ages for the galaxies
analyzed here. 
That leads to think that these three  objects present different 
phases of a single phenomenon, in spite of  the strong bar in 
NGC\,1365. These conclusions have to be further checked with
more elaborated models and for a larger sample of galaxies.

\section*{Acknowledgements}

Nelson Vera-Villamizar thanks a studentship of the Brazilian 
Foundation CNPq. This research is partially supported by the 
Mexican Foundation CONACYT under the grant No. 28507-E
and by the Brazilian institutions CNPq and CAPES under the
program PRONEX grant No. 76.97.1003.00.

\newpage


\begin{figure}
\vspace{8.0 cm}
\begin{tabular}{cc}
\end{tabular}
\begin{picture}(0,0)
\put(0,200){(a)}
\put(245,200){(b)}
\end{picture}
\caption{NGC\,1365 disk subtracted and rectified to face-on 
images according to procedure described in section 2.1. (a) 
$g$ image, (b) $i$ image.}
\label{ima1365_g_i}
\end{figure}

\begin{figure}
\vspace{8.0 cm}
\begin{tabular}{cc}
\includegraphics{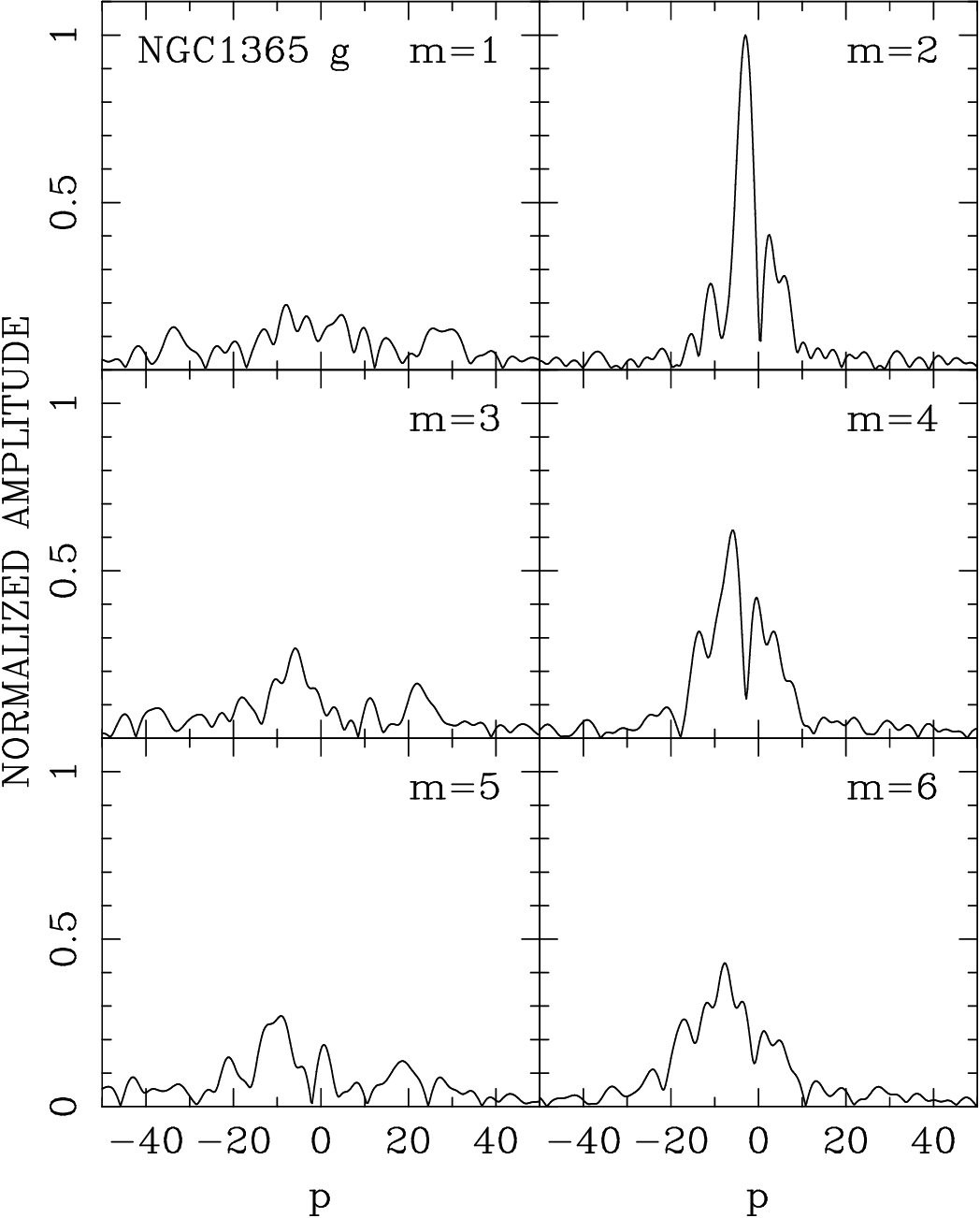}&
\includegraphics{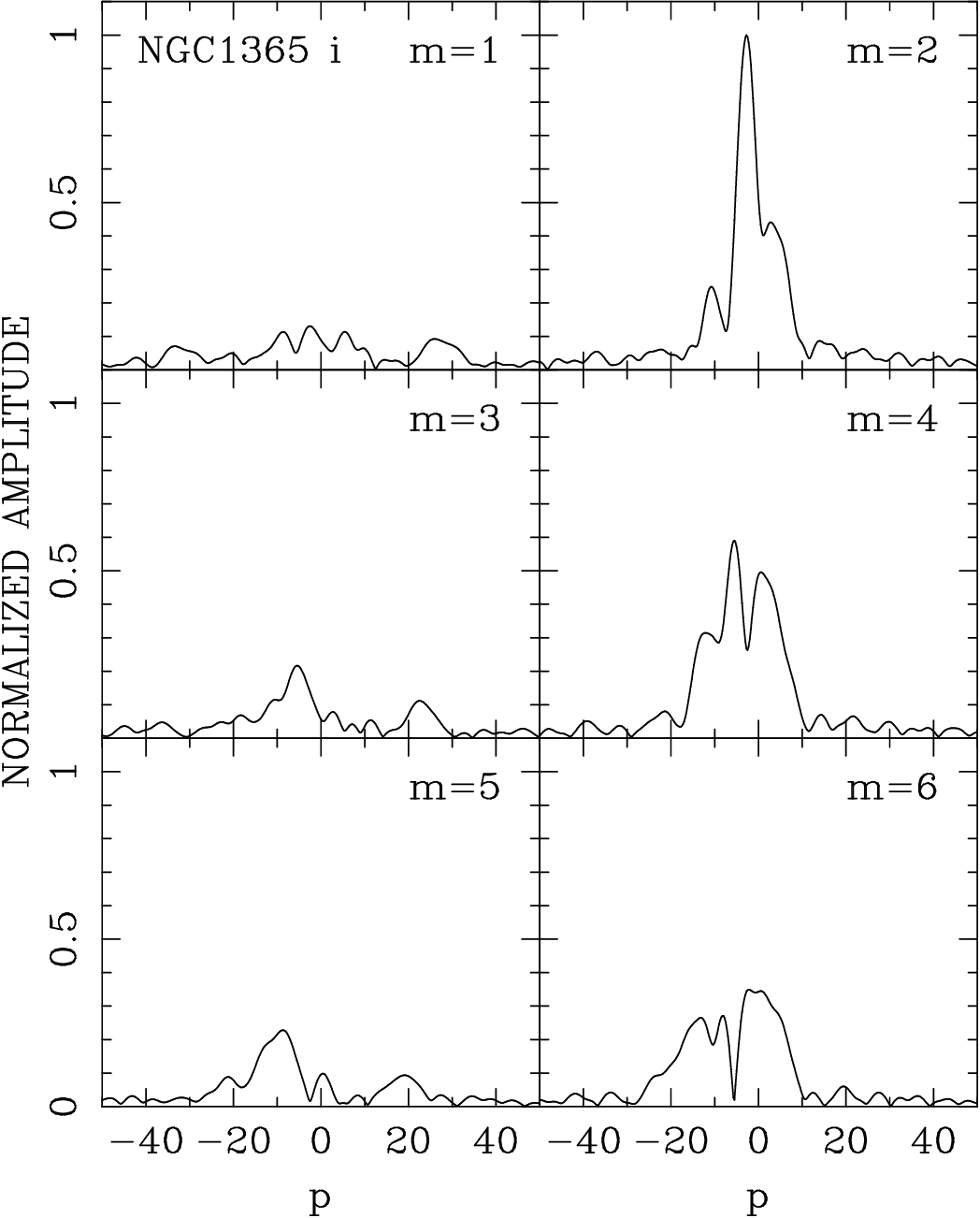} \cr
\end{tabular}
\caption{Two-dimensional Fourier spectra coefficients $A(p,m)$ of 
NGC\,1365, obtained as described in section 2.4.2. Left, $g$ image,
and right $i$ image. The $p$ variable is related to the pitch angle 
$\alpha$ by $p=-\frac{m}{\tan \alpha}$.}
\label{fouspec1365}
\end{figure}

\clearpage

\begin{figure}[p]
\vspace{9.7 cm}
\caption{NGC\,1365 $g$ band rectified image in $\log$ r versus 
$\theta$ diagram. The horizontal line shows the CR loci at 
13 kpc. According to \cite{eem92}(Hereinafter EEM92).}
\label{rectifica1365}
\end{figure}

\clearpage

\begin{figure}
\vspace{9.7 cm}
\includegraphics{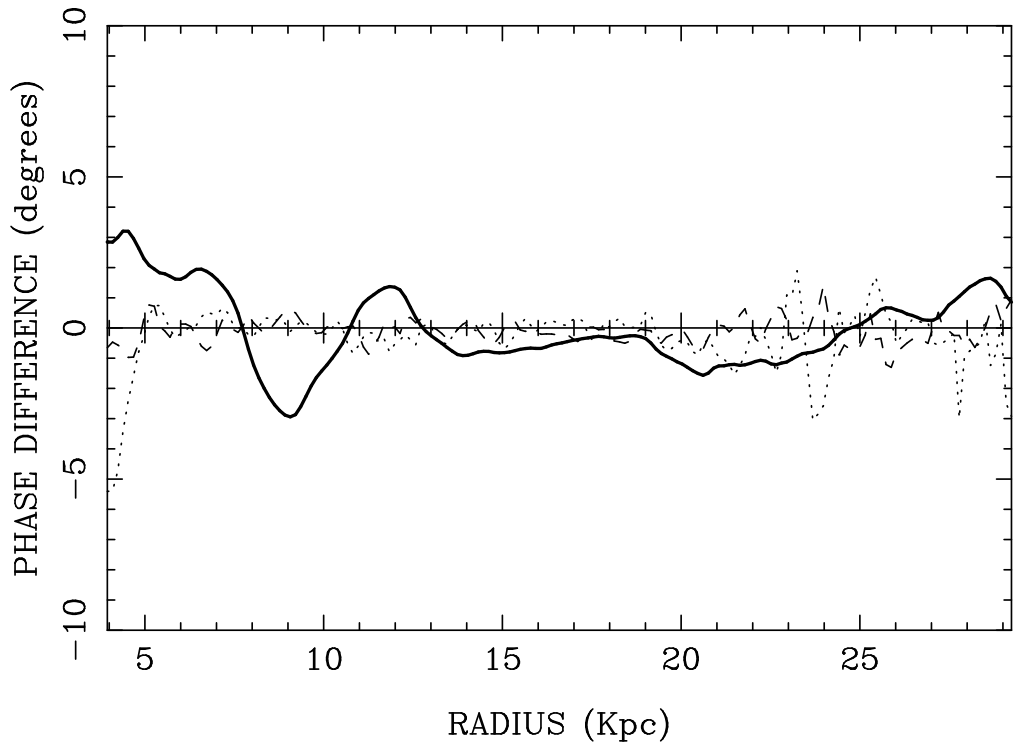}
\caption{Azimuthal Profile Phase Difference (APPD) 
$[\Theta_{2g}-\Theta_{2i}]$ for NGC\,1365 obtained 
from original images (solid line). Broken line shows $g$ noise and dotted 
line shows $i$ noise.}
\label{phase+error_pure1365}
\end{figure}

\begin{figure}
\vspace{9.7 cm}
\includegraphics{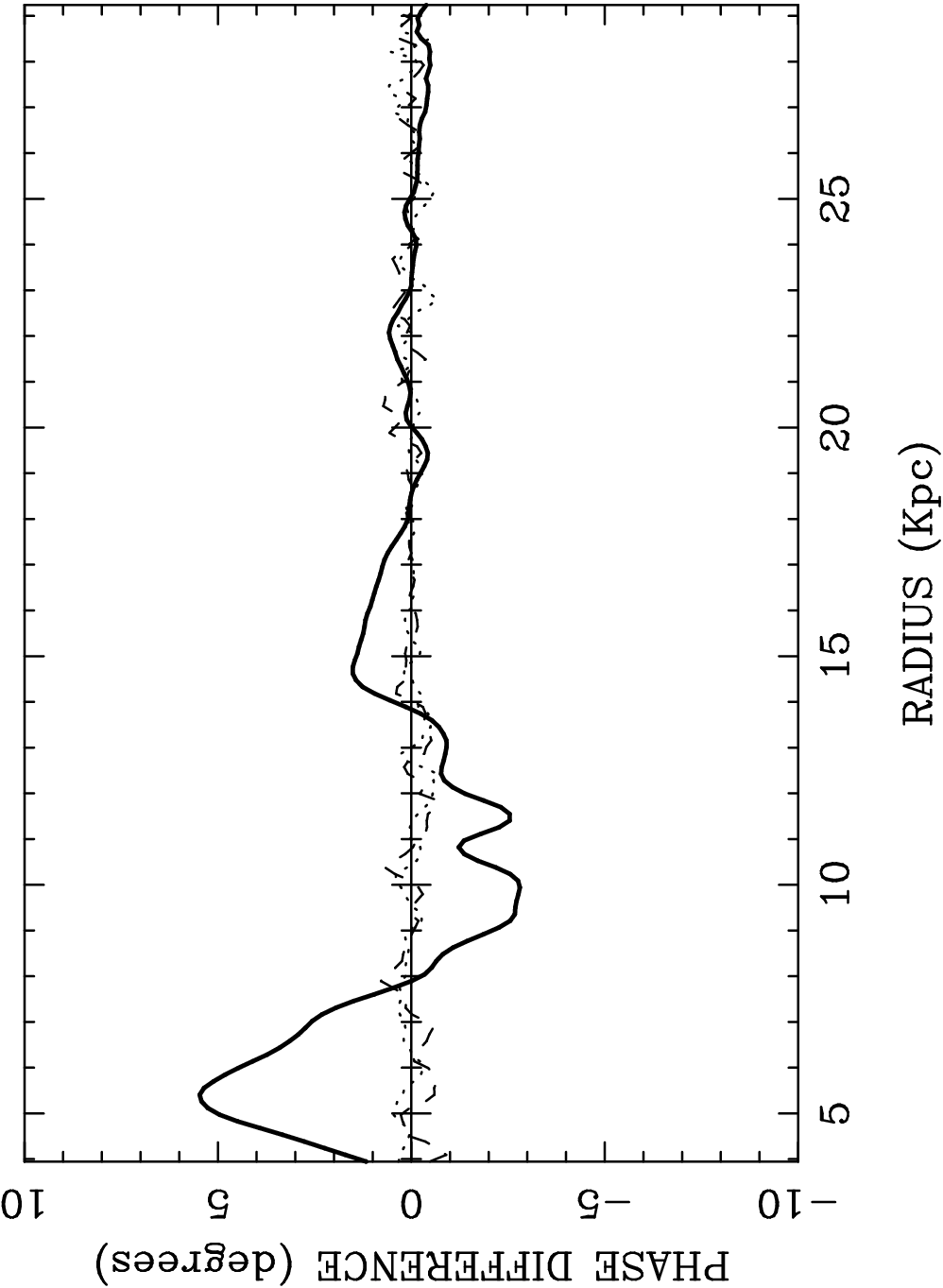}
\caption{APPD $[\Theta_{2g}-\Theta_{2i}]$ for NGC\,1365 
obtained from images symmetrized according to EEM92
Method (solid line). Broken line shows the 
noise in $g$ color and dotted line shows that in $i$ color. 
See sections 2.4.1 and 2.3.}
\label{phase+error_eem1365}
\end{figure}

\begin{figure}
\vspace{9.7 cm}
\includegraphics{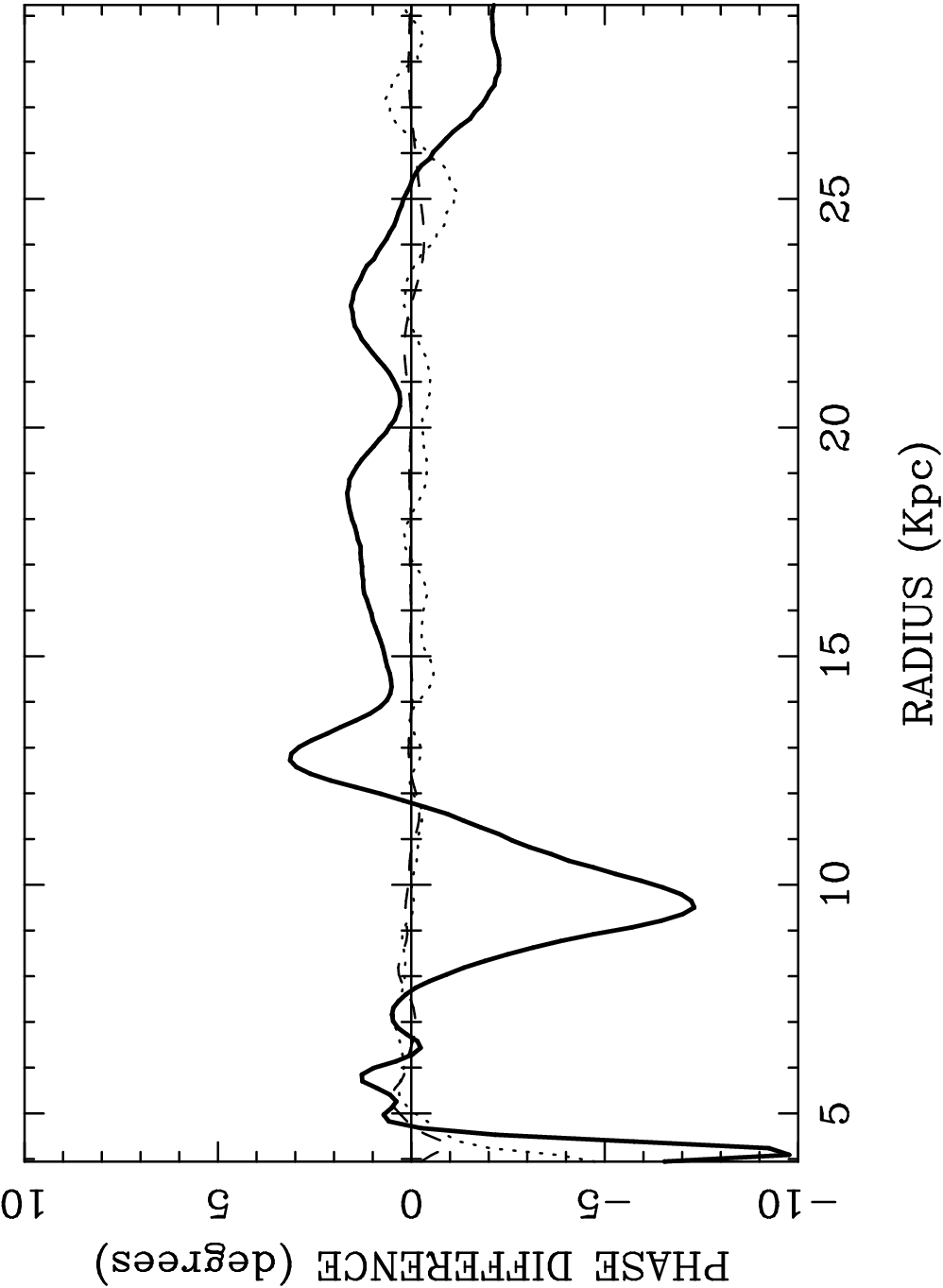}
\caption{APPD $[\Theta_{2g}-\Theta_{2i}]$ for NGC\,1365 
obtained from the Fourier m=2 modes (solid line). Broken line shows $g$ 
noise and dotted line shows $i$ noise. See sections 2.4.2 and 2.3.}
\label{phase+error_four1365}
\end{figure}

\begin{figure}
\vspace{10.2 cm}
\includegraphics{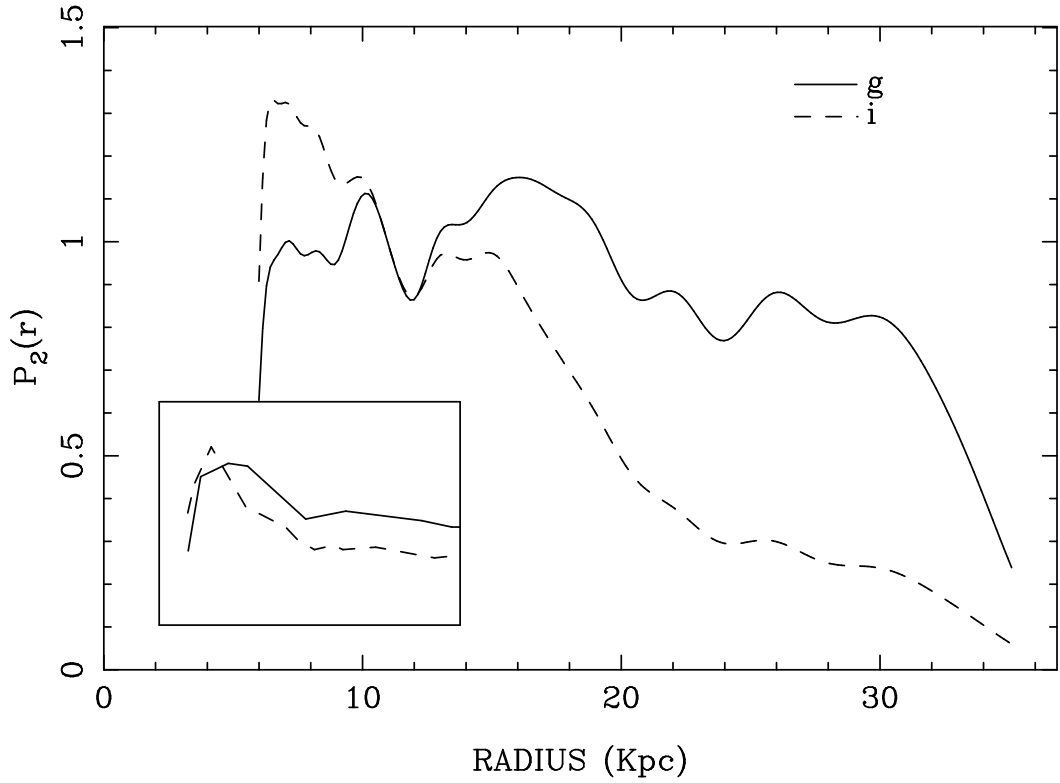}
\caption{Comparison of Fourier radial density functions in blue 
$P_{2g}$ and infrared $P_{2i}$ (see section 2.4.2) for 
NGC\,1365. Inset shows SDW and gas Fourier radial density 
functions for \cite{juncom96} model E after 1200 Myr.}
\label{smr1365}
\end{figure}

\clearpage

\begin{figure}
\vspace{8.0 cm}
\begin{tabular}{cc}
\end{tabular}
\begin{picture}(0,0)
\put(0,200){(a)}
\put(245,200){(b)}
\end{picture}
\caption{NGC\,1566 disk subtracted and rectified to face-on 
images according to procedure described in section 2.1. (a) 
$g$ image, (b) $i$ image.}
\label{ima1566_g_i}
\end{figure}

\begin{figure}
\vspace{8.0 cm}
\begin{tabular}{cc}
\includegraphics{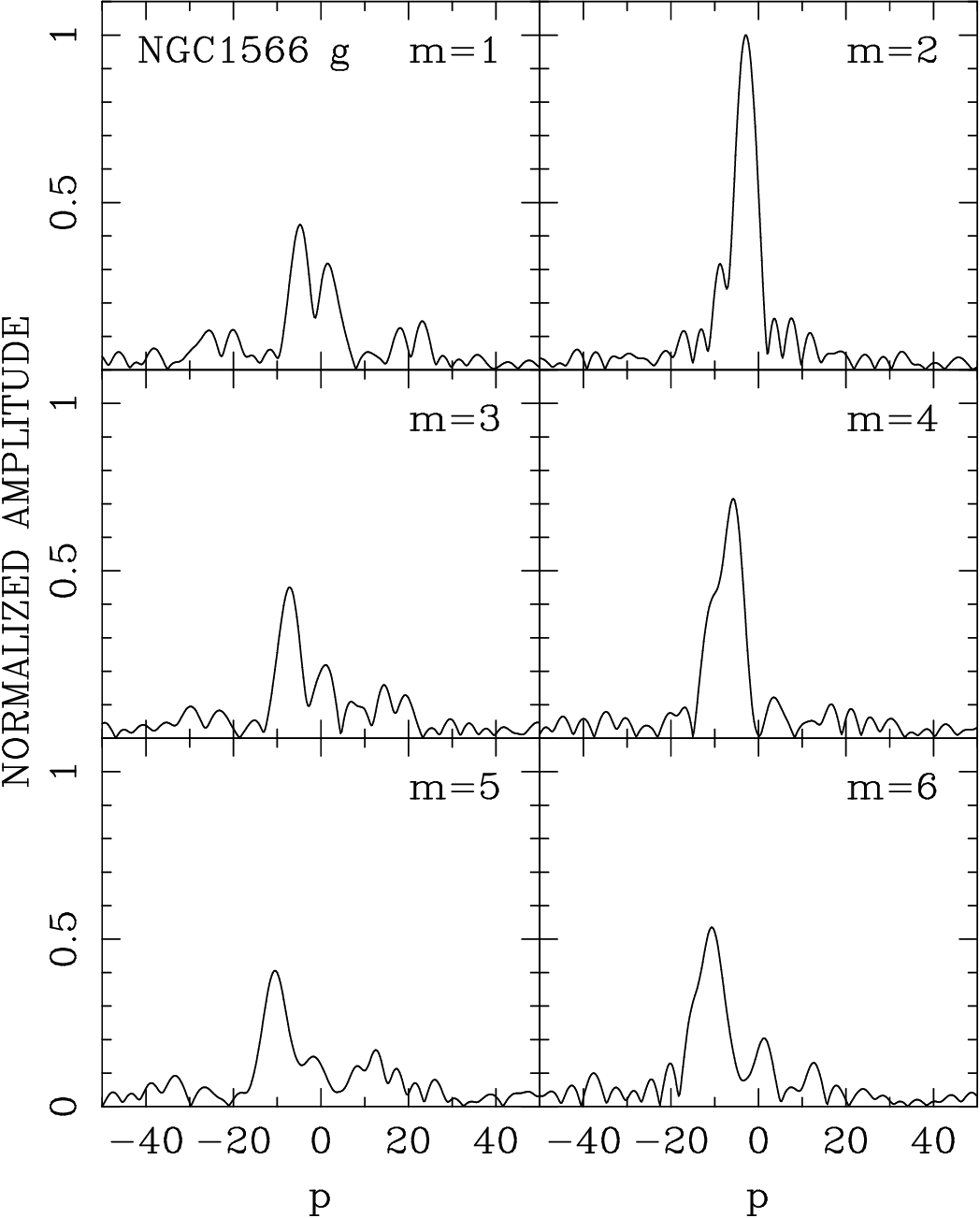}&
\includegraphics{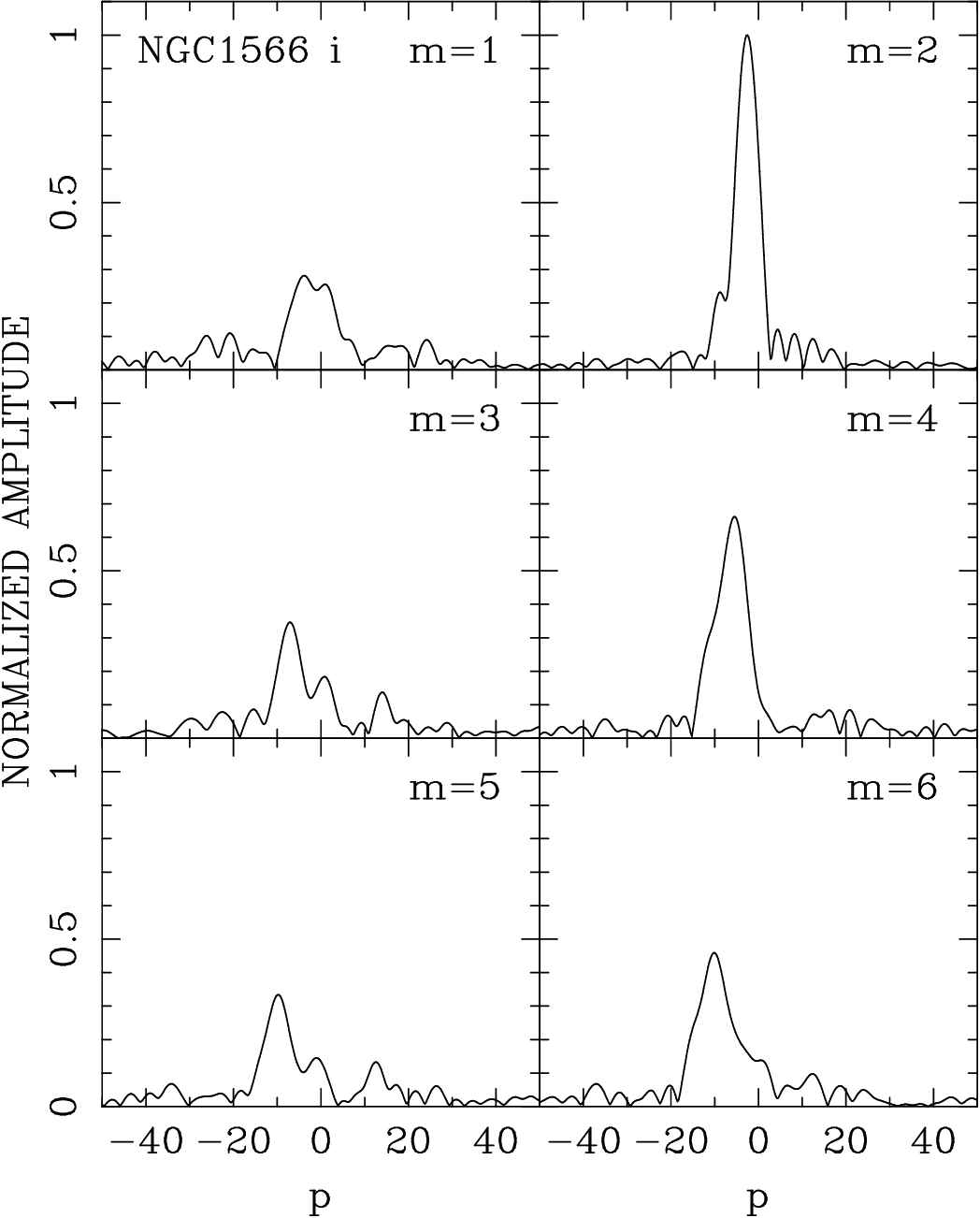} \cr
\end{tabular}
\caption{Two-dimensional Fourier spectra coefficients $A(p,m)$ of 
NGC\,1566, obtained as described in section 2.4.2. Left, $g$ image,
and right $i$ image. The $p$ variable 
is related to the pitch angle $\alpha$ by $p=-\frac{m}{\tan \alpha}$.}
\label{fouspec1566}
\end{figure}

\clearpage

\begin{figure}
\vspace{9.7 cm}
\caption{NGC\,1566 $g$ band rectified image in $\log$ r versus 
$\theta$ diagram. The horizontal line shows the CR loci at 
9.4 kpc. According to EEM92.}
\label{rectifica1566}
\end{figure}

\clearpage

\begin{figure}
\vspace{9.7 cm}
\includegraphics{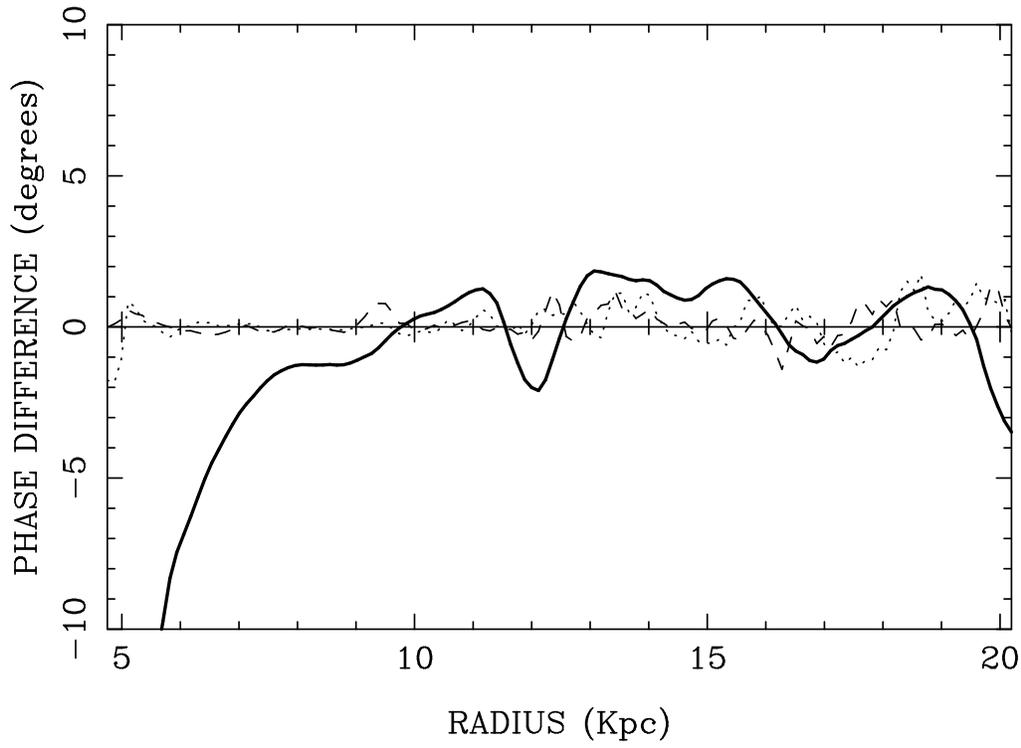}
\caption{APPD $[\Theta_{2g}-\Theta_{2i}]$ for NGC\,1566 
obtained from original images (solid line). Broken line shows $g$ noise 
and dotted line shows $i$ noise.}
\label{phase+error_pure1566}
\end{figure}

\begin{figure}
\vspace{9.7 cm}
\includegraphics{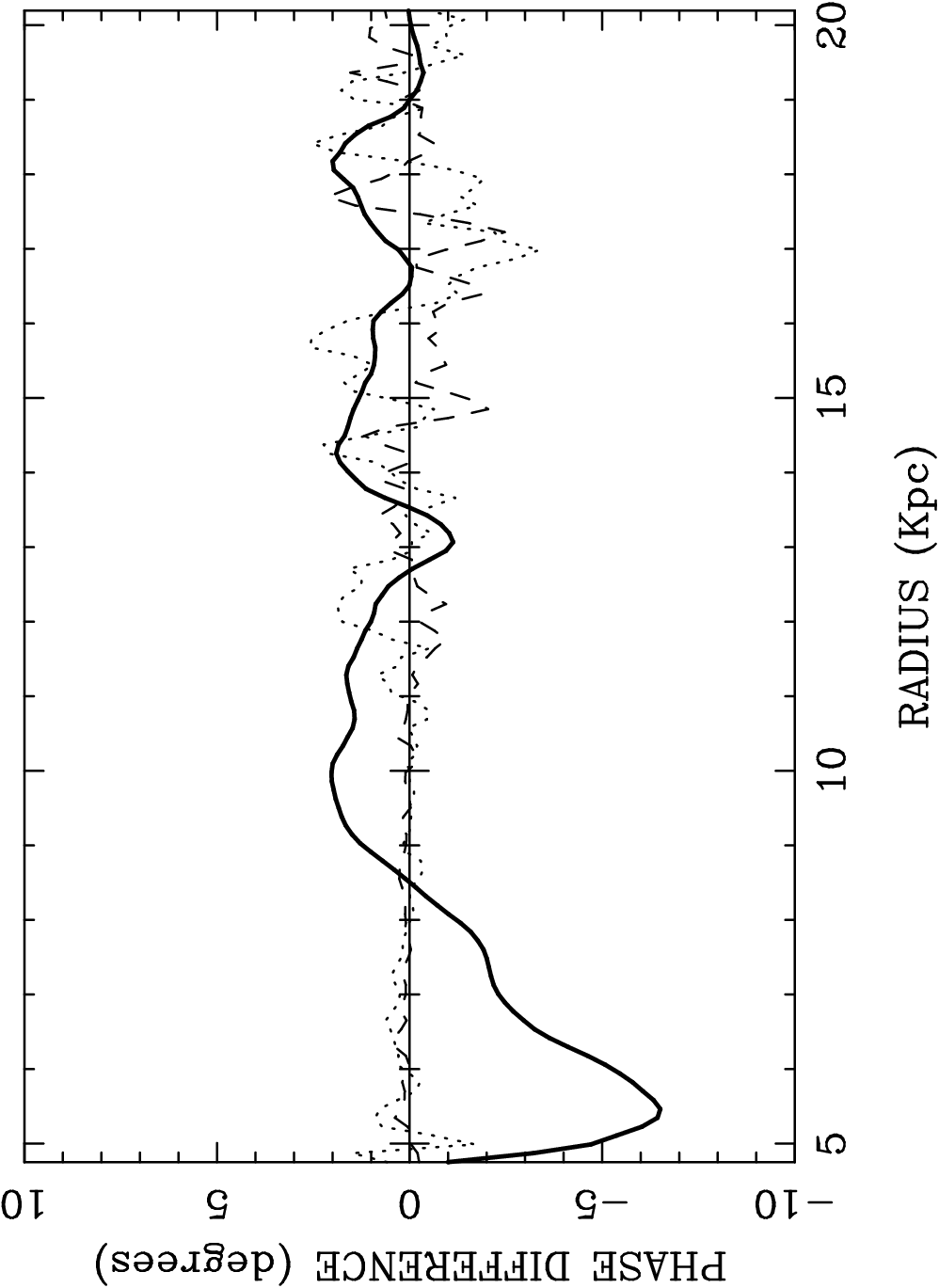}
\caption{APPD $[\Theta_{2g}-\Theta_{2i}]$ for NGC\,1566 
obtained from images symmetrized according to EEM92
Method (solid line). Broken line shows the 
noise in $g$ color and dotted line shows that in $i$ 
color. See sections 2.4.1 and 2.3.}
\label{phase+error_eem1566}
\end{figure}

\clearpage

\begin{figure}
\vspace{9.7 cm}
\includegraphics{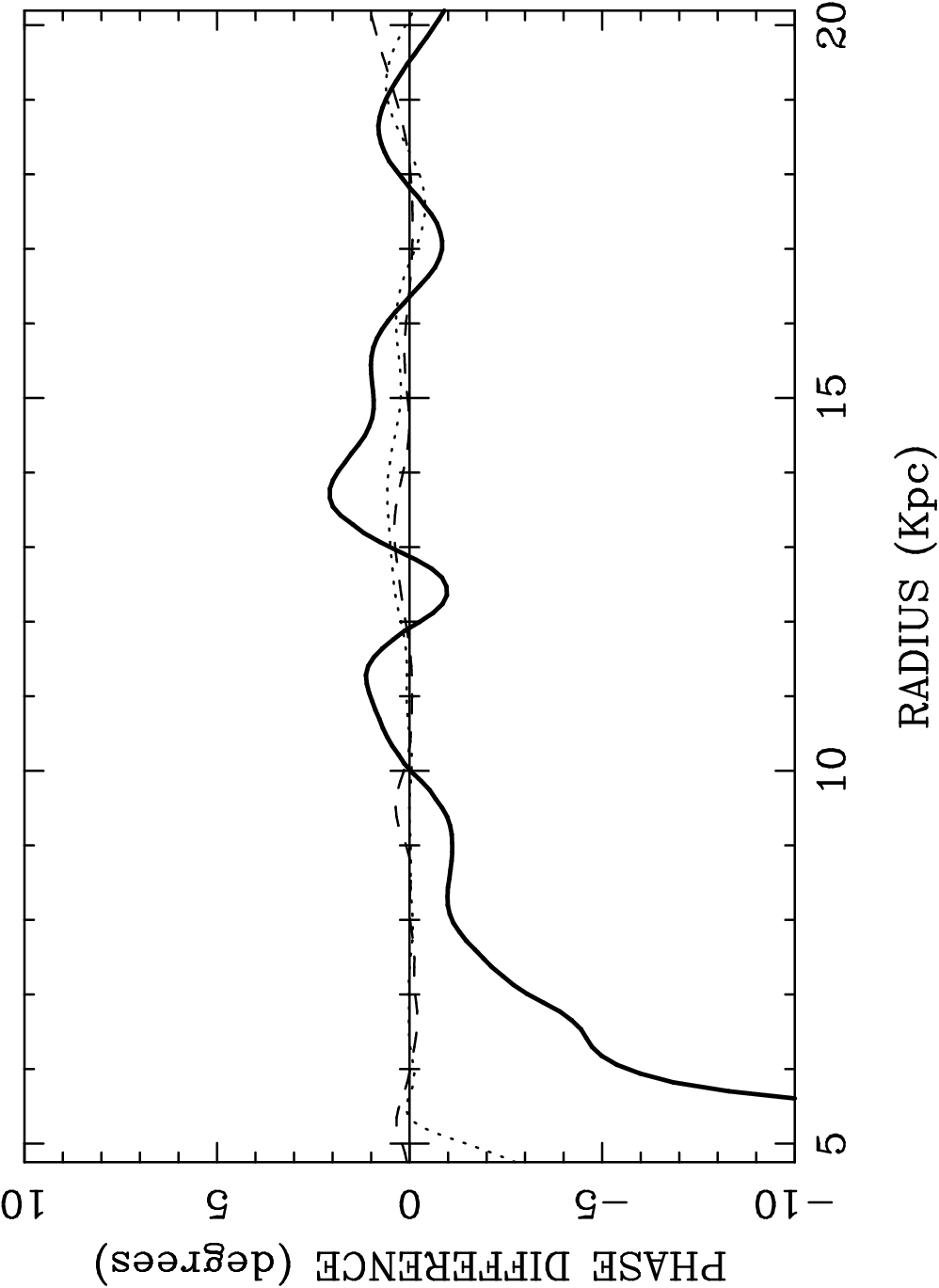}
\caption{APPD $[\Theta_{2g}-\Theta_{2i}]$ for NGC\,1566 
obtained from the Fourier m=2 modes (solid line). Broken line shows $g$ 
noise and dotted line shows $i$ noise. See section 2.4.2.}
\label{phase+error_four1566}
\end{figure}

\begin{figure}
\vspace{9.8 cm}
\includegraphics{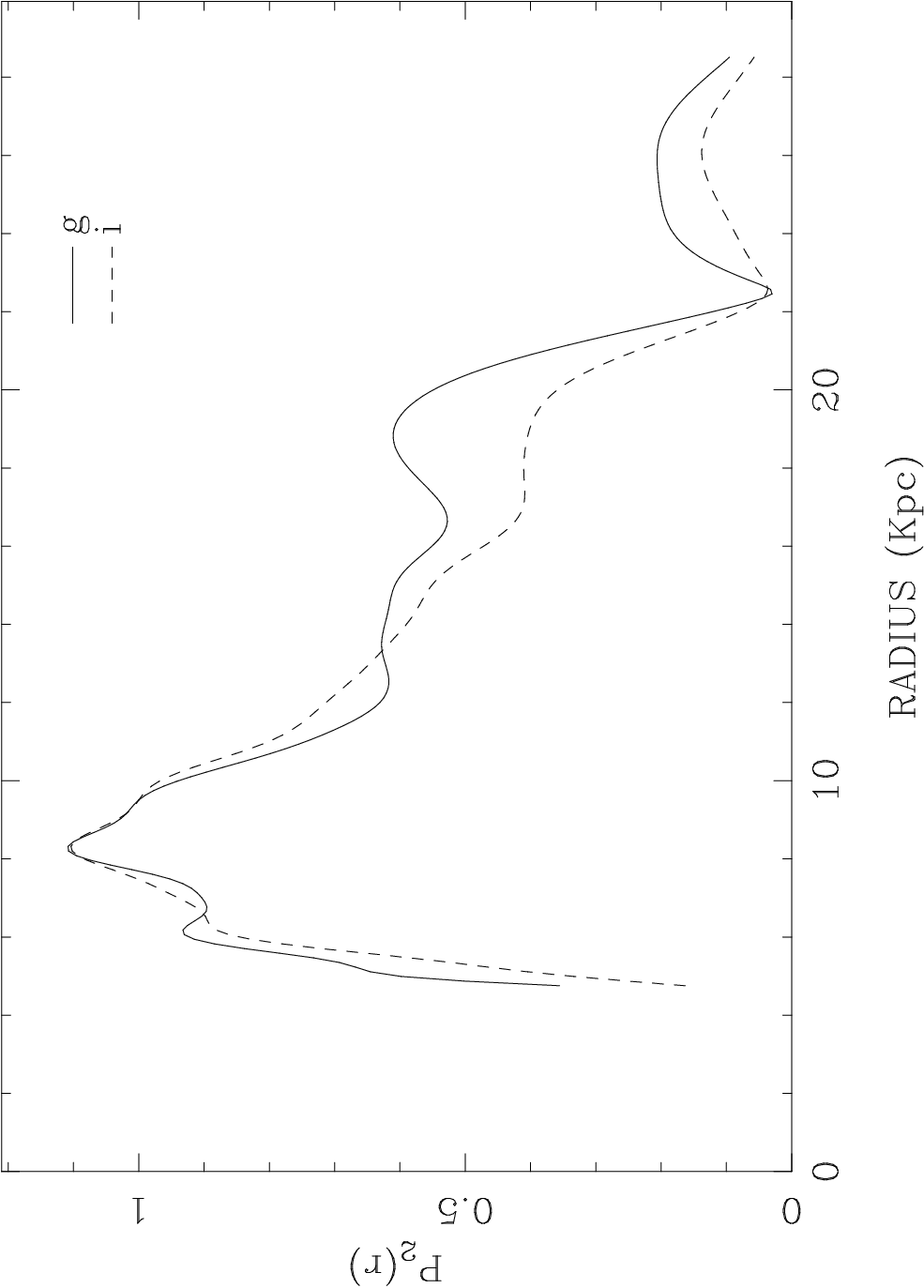}
\caption{Comparison of Fourier radial density functions in blue 
$P_{2g}$ and infrared $P_{2i}$ (see section 2.4.2) for 
NGC\,1566.}
\label{smr1566}
\end{figure}

\begin{figure}
\vspace{9.8 cm}
\includegraphics{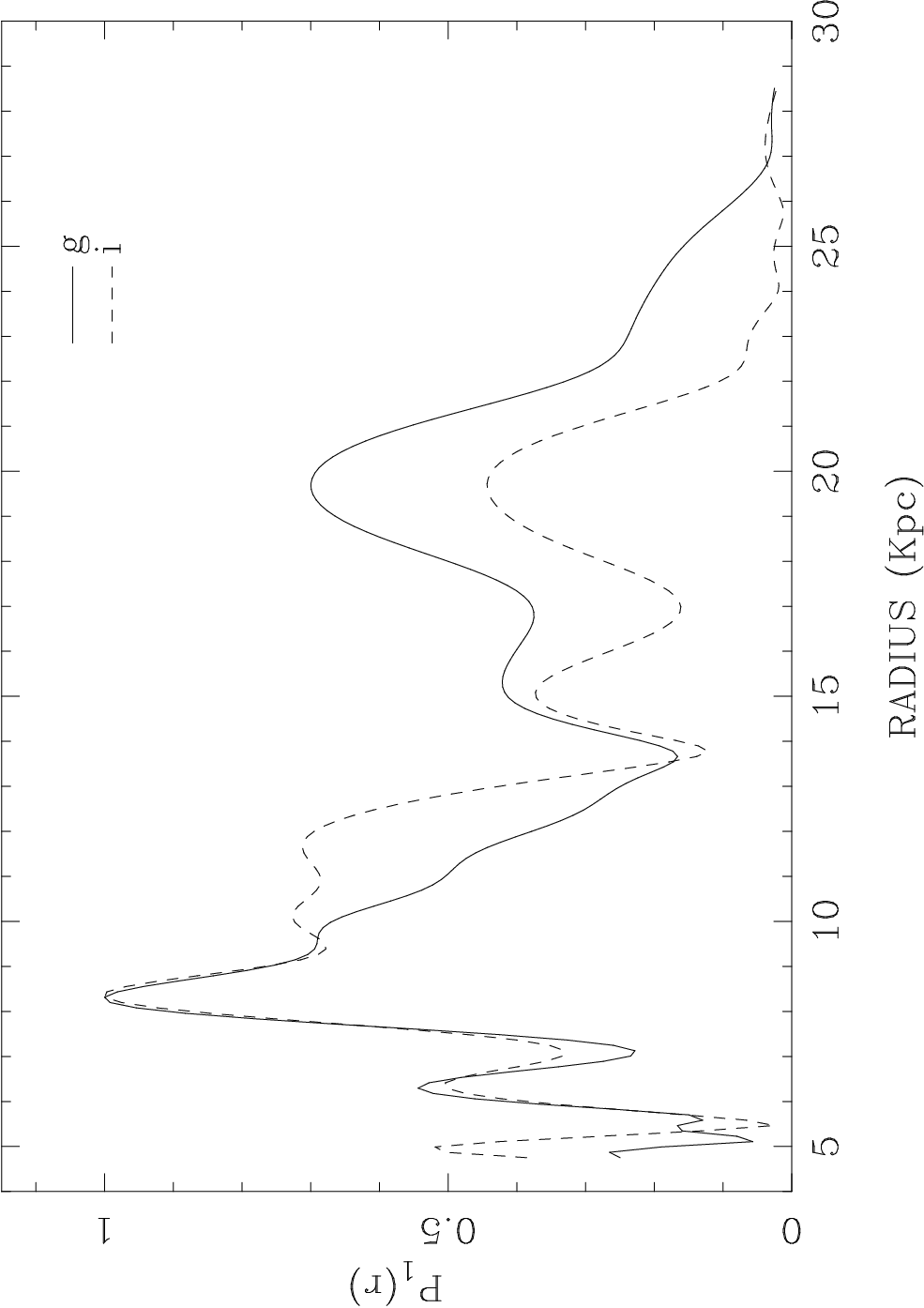}
\caption{Comparison of Fourier radial density functions in blue 
$P_{1g}(r)$ and infrared $P_{1i}(r)$ (see section 2.4.2) for 
NGC\,1566.}
\label{sm=1r1566}
\end{figure}

\clearpage

\begin{figure}
\vspace{8.0 cm}
\begin{tabular}{cc}
\end{tabular}
\begin{picture}(0,0)
\put(0,200){(a)}
\put(245,200){(b)}
\end{picture}
\caption{NGC\,2997 disk subtracted and rectified to face-on 
images according to procedure described in section 2.1. (a) 
$g$ image, (b) $i$ image.}
\label{ima2997_g_i}
\end{figure}

\begin{figure}
\vspace{8.0 cm}
\begin{tabular}{cc}
\includegraphics{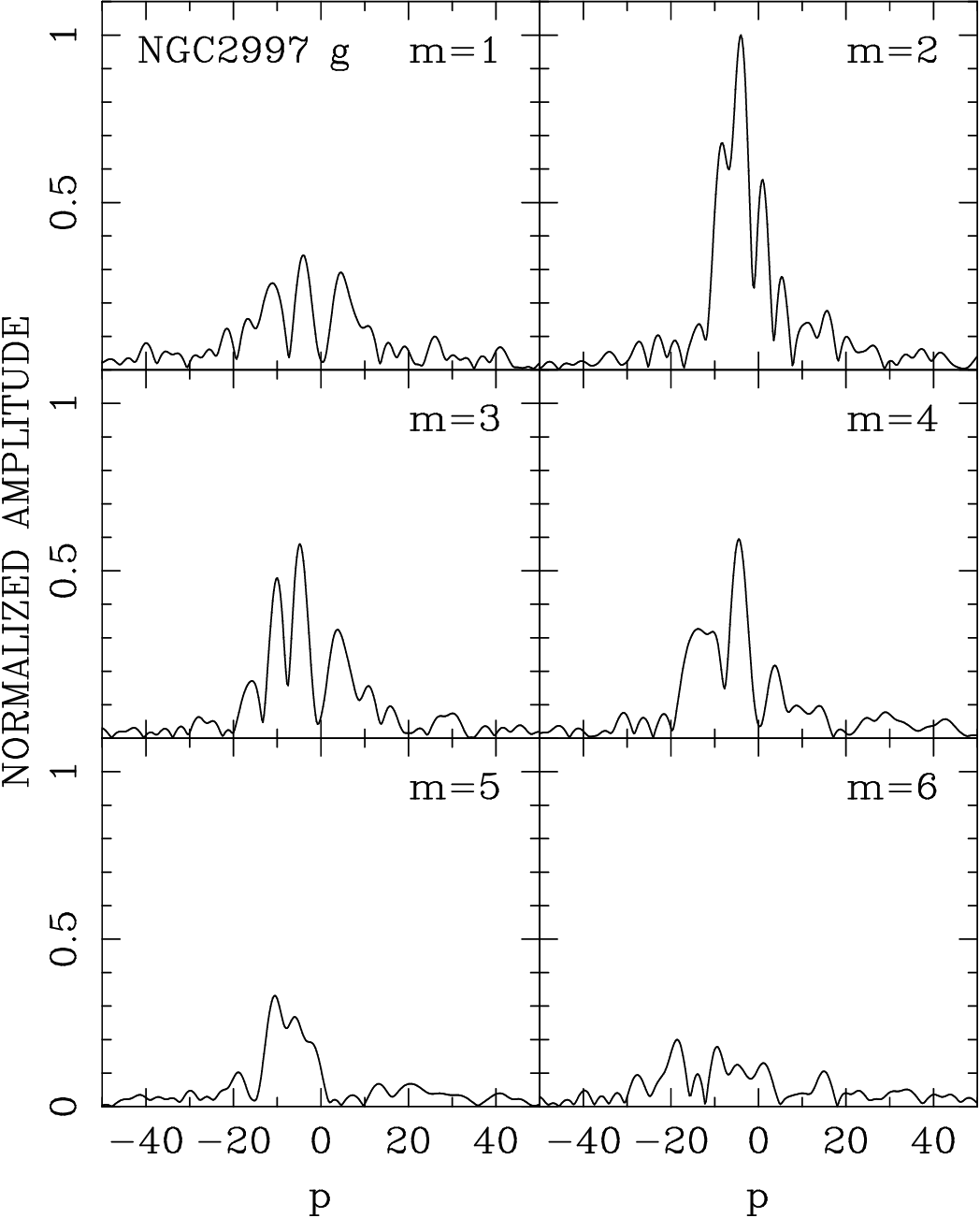}&
\includegraphics{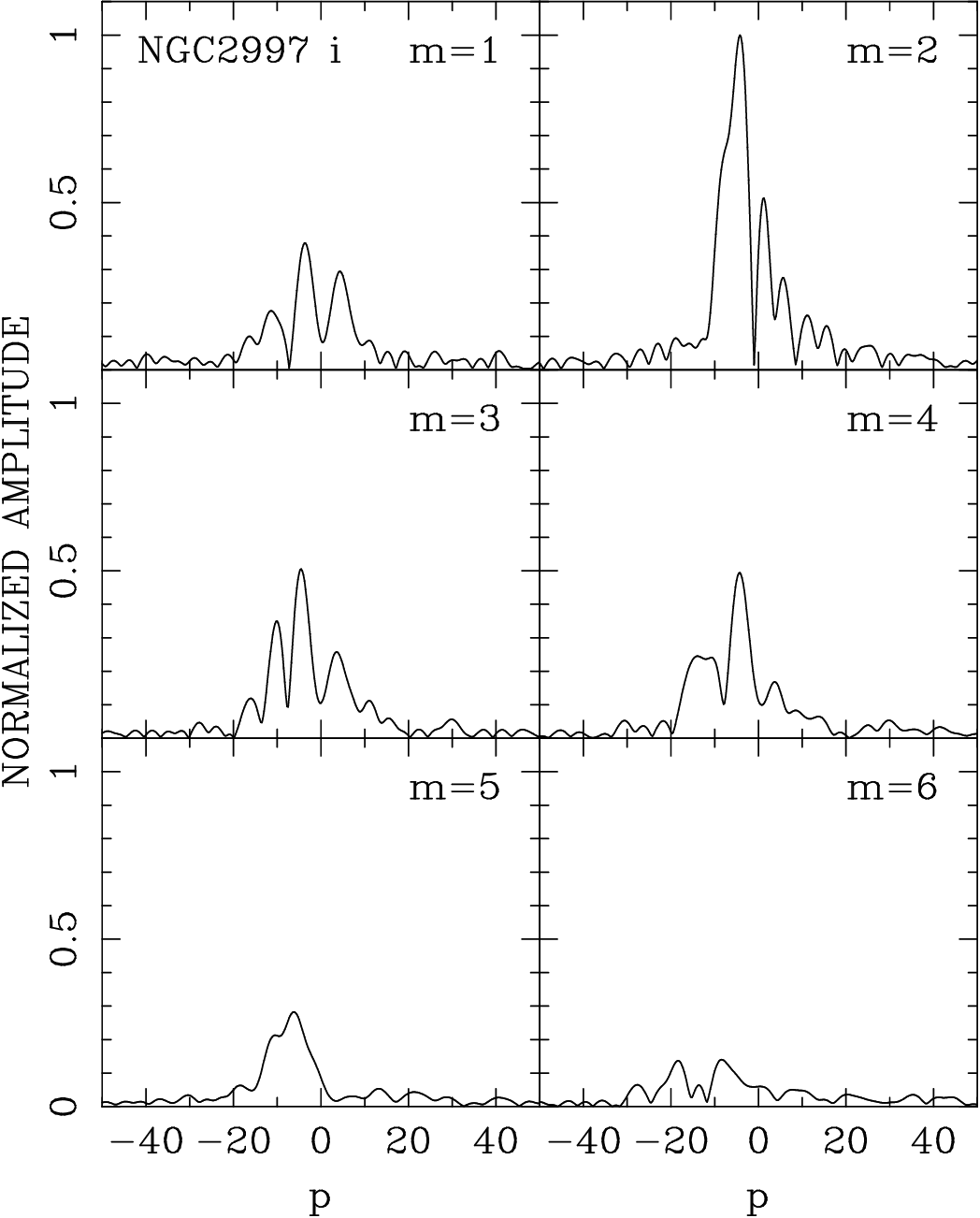} \cr
\end{tabular}
\caption{Two-dimensional Fourier spectra coefficients $A(p,m)$ of 
NGC\,2997, obtained as described in section 2.4.2. Left, $g$ image,
and right $i$ image. The $p$ variable 
is related to the pitch angle $\alpha$ by $p=-\frac{m}{\tan \alpha}$.}
\label{fouspec2997}
\end{figure}

\clearpage

\begin{figure}
\vspace{9.7 cm}
\caption{NGC\,2997 $g$ band symmetrized EEM92 S2 and rectified  
image in $\log$ R versus $\theta$ diagram. The horizontal 
line shows the CR loci at 7\,kpc. According to EEM92.}
\label{rectifica2997}
\end{figure}

\begin{figure}
\vspace{9.7 cm}
\caption{NGC\,2997 S2 symmetrized image according to EEM92
Method.}
\label{s2innerouter2997}
\end{figure}

\clearpage

\begin{figure}
\vspace{9.7 cm}
\includegraphics{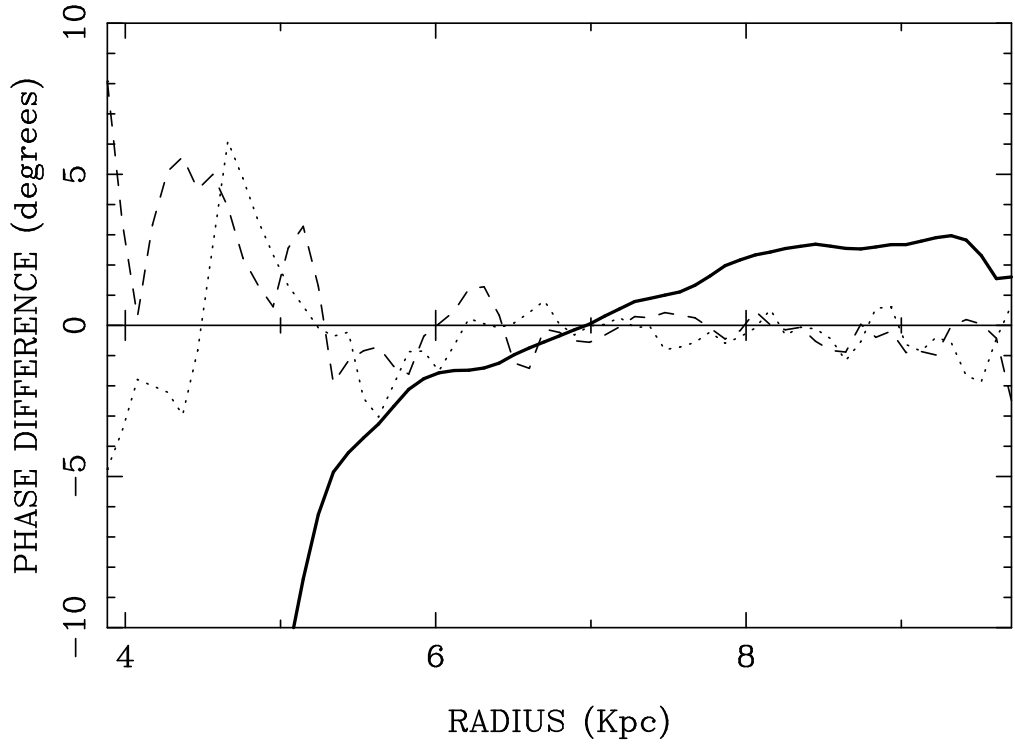}
\caption{APPD $[\Theta_{2g}-\Theta_{2i}]$ for NGC\,2997 
obtained from images symmetrized according to EEM92
Method, inner arms between 
$4<r<11$\,kpc (solid line). Broken line shows the noise in $g$ color 
and dotted line shows that in $i$ color.}
\label{phase+error_eem2997in}
\end{figure}

\begin{figure}
\vspace{9.7 cm}
\includegraphics{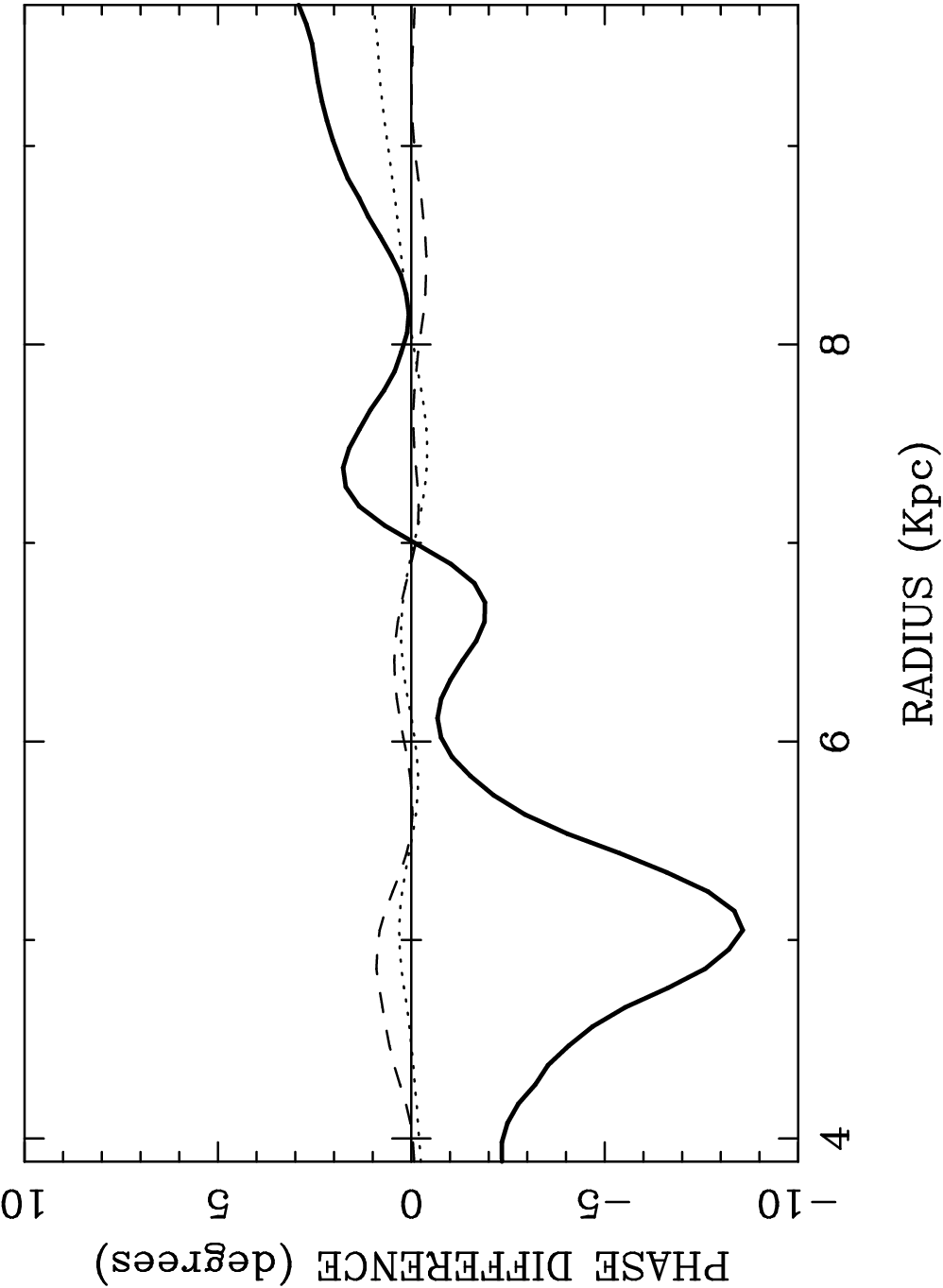}
\caption{APPD $[\Theta_{2g}-\Theta_{2i}]$ for NGC\,2997
obtained from the Fourier m=2 modes (solid line). Broken line shows $g$ 
noise and dotted line shows $i$ noise. See section 2.4.2.}
\label{phase+error_four2997}
\end{figure}

\begin{figure}
\vspace{9.7 cm}
\includegraphics{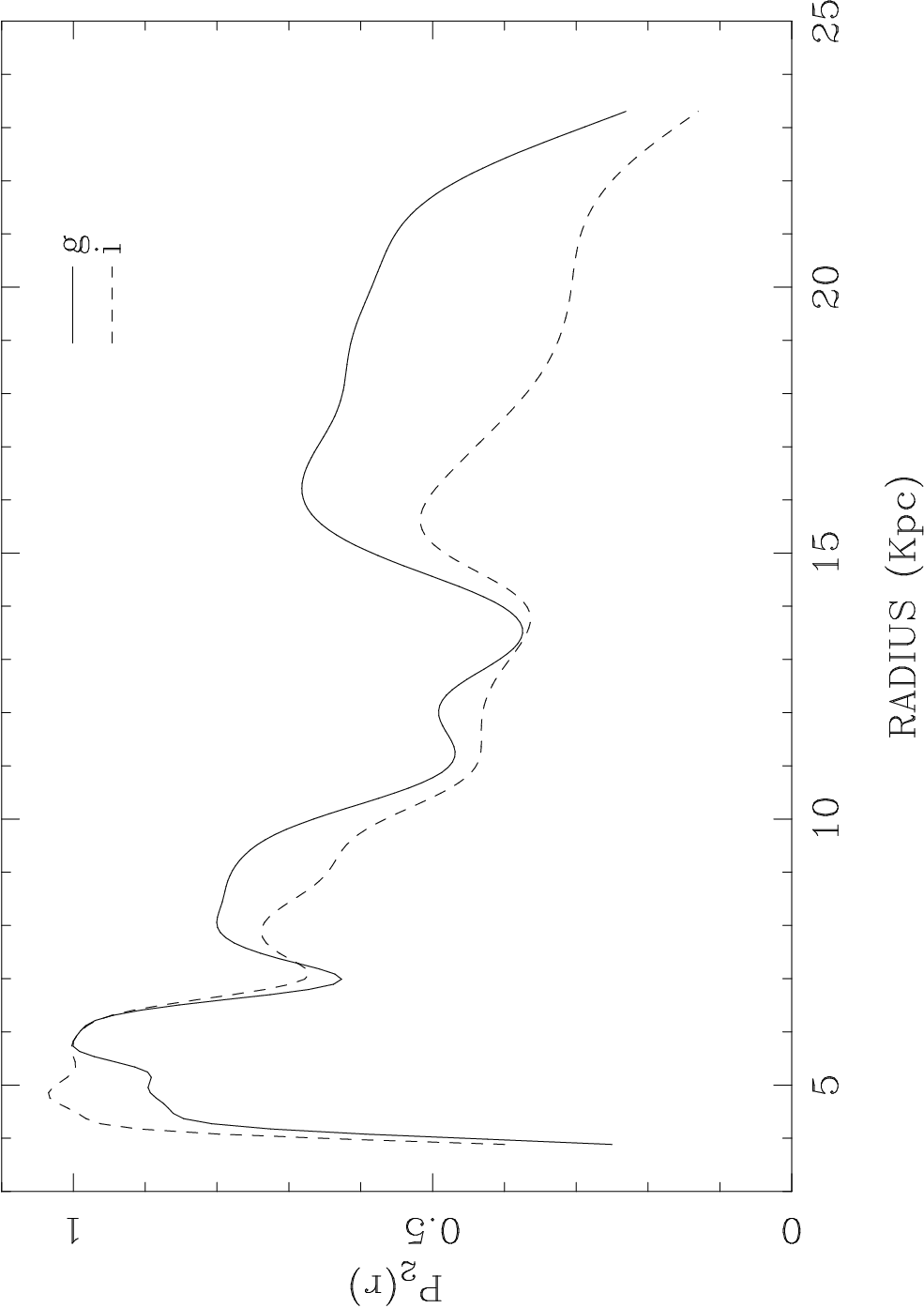}
\caption{Comparison of Fourier radial density functions in blue 
$P_{2g}$ and infrared $P_{2i}$ (see section 2.4.2) for 
NGC\,2997.}
\label{smr2997}
\end{figure}

\clearpage

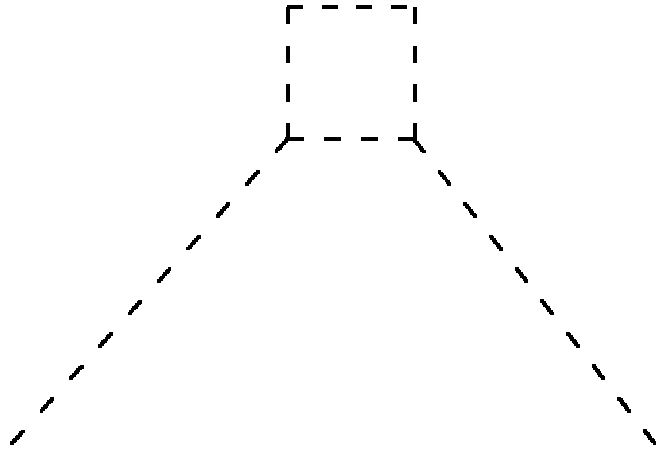
\begin{figure}
\vspace{20 cm}
\begin{picture}(0,0)
\linethickness{1pt}
\dashline[0]{7}(220,448)(265,448)
\dashline[0]{7}(220,448)(220,398)
\dashline[0]{7}(268,448)(268,398)
\dashline[0]{7}(220,398)(268,398)
\dashline[0]{7}(220,398)(116,283)
\dashline[0]{7}(268,398)(359,283)
\end{picture}
\caption{NGC\,2997 S3 symmetrized image according to EEM92
Method. Curves drawn at the 3-arm beginning points to the
leading fronts of star formation. The blow-up shows details 
of one of the three fronts. See section 5.1 paragraph 3.}
\label{s3_2997}
\end{figure}

\clearpage

\begin{figure}
\vspace{9.7 cm}
\includegraphics{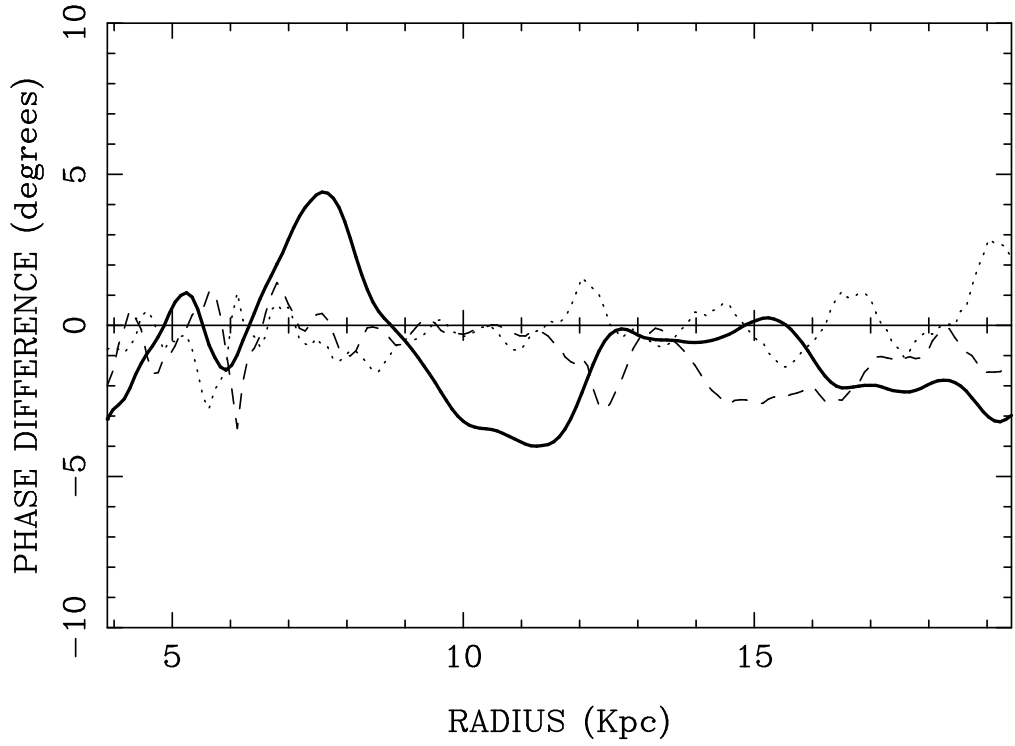}
\caption{APPD $[\Theta_{3g}-\Theta_{3i}]$ for NGC\,2997 
obtained from m=3 Fourier modes (solid line). This Fourier analysis was
performed in S3 symmetrized images according to EEM92
Method. Broken line shows the noise in $g$ 
color and dotted line shows that in $i$ color.}
\label{phase+error_eem2997s3}
\end{figure}

\clearpage

\begin{deluxetable}{lccccccc}
\tablewidth{40pc}
\tablecaption{Galaxy parameters. \label{galaxies}}
\tablehead{
\colhead{Galaxy}	& \colhead{Type}	&
\colhead{RA}		& \colhead{Dec}		& 
\colhead{i\degr} 		& \colhead{PA\degr}		& 
\colhead{v\,(km/s)}		& \colhead{Conv\,(kpc/\arcsec)}	}
\startdata
NGC\,1365 & SBb(s)b & 03h33m36.4s  & -36d08m25s & 40 & 220 & 1636 & 0.1074 \\
NGC\,1566 & SAB(rs)bc & 04h20m00.6s & -54d56m17s & 37 & 60 & 1496 & 0.0967 \\
NGC\,2997 & SA(s)c & 09h45m38.6s & -31d11m25s & 40 & 80 & 1087 & 0.073 \\
\enddata
\end{deluxetable}

\begin{deluxetable}{lccccc}
\tablewidth{27pc}
\tablecaption{Resonances mean value. h=75\,Km/s/Mpc\label{resonances}}
\tablehead{
\colhead{Galaxy}       & \colhead{CR}   &
\colhead{ILR}          & \colhead{OLR} & \colhead{m} &
\colhead{$\Omega_{CR}(km/s/kpc)$}	}
\startdata
NGC1365\tablenotemark{a} & 12.1$\pm$0.9 & 0.6 & 18.3 & 2 & 25.0$\pm$1.9 \\
NGC1566\tablenotemark{b} & 9.4$\pm$0.4 & 3.2 & 14.6 & 2 & 12.2$\pm0.9$ \\
  & 7.1\tablenotemark{c} & \nodata & 14.8 & 1 & 16.6 \\
NGC2997\tablenotemark{d} & 7.0 & 2.7 & 9.7 & 2 & 17.6 \\
  & 8.7 & 6.0 & 10.9 & 3 & 12.7 \\
\tablenotetext{a}{The rotation curve was taken from \cite{per_sal95} and \cite{Jors95}.}
\tablenotetext{b}{The rotation curve was taken from \cite{per_sal95}.}
\tablenotetext{c}{One-arm CR was estimated from Log r vs
$\Theta$ diagram of the antisymmetric images $g$ and $i$ according to
Elmegreen, Elmegreen \& Montenegro method. The S/N is low and the value
should be taken with ware.}
\tablenotetext{d}{The rotation curve was taken from \cite{sperandio95}.}
\enddata
\end{deluxetable}

\end{document}